\documentclass{jpp}
\usepackage{graphicx}
\usepackage{afterpage}
\usepackage[utf8]{inputenc}
\usepackage[T1]{fontenc}
\usepackage{amsmath}
\usepackage{cleveref}
\usepackage{cancel}
\usepackage{xcolor} 
\usepackage{array}

\shorttitle{Elasticity of tangled magnetic fields}
\shortauthor{D. N. Hosking, A. A. Schekochihin and S. A. Balbus}

\title{Elasticity of tangled magnetic fields}

\author{D. N. Hosking\aff{1,}\aff{3}
  \corresp{\email{david.hosking@physics.ox.ac.uk}}, A. A. Schekochihin\aff{2,}\aff{3}, \and S. A. Balbus\aff{1,}\aff{4}}

\affiliation{\aff{1}Oxford Astrophysics, Denys Wilkinson Building, Keble Road, Oxford OX1 3RH, UK
\aff{2}Rudolf Peierls Centre for Theoretical Physics, Clarendon Laboratory, Parks Road, Oxford, OX1 3PU, UK
\aff{3}Merton College, Merton Street, Oxford, OX1 4JD, UK
\aff{4}New College, Holywell Street, Oxford OX1 3BN, UK}

\begin{document}

\maketitle

\begin{abstract}

The fundamental difference between incompressible ideal magnetohydrodynamics and the dynamics of a non-conducting fluid is that magnetic fields exert a tension force that opposes their bending; magnetic fields behave like elastic strings threading the fluid. It is natural, therefore, to expect that a magnetic field tangled at small length scales should resist a large-scale shear in an elastic way, much as a ball of tangled elastic strings responds elastically to an impulse. Furthermore, a tangled field should support the propagation of `magnetoelastic waves', the isotropic analogue of Alfvén waves on a straight magnetic field. Here, we study magnetoelasticity in the idealised context of an equilibrium tangled field configuration. In contrast to previous treatments, we explicitly account for intermittency of the Maxwell stress, and show that this intermittency necessarily decreases the frequency of magnetoelastic waves in a stable field configuration. We develop a mean-field formalism to describe magnetoelastic behaviour, retaining leading-order corrections due to the coupling of large- and small-scale motions, and solve the initial-value problem for viscous fluids subjected to a large-scale shear, showing that the development of small-scale motions results in anomalous viscous damping of large-scale waves. Finally, we test these analytic predictions using numerical simulations of standing waves on tangled, linear force-free magnetic-field equilibria.

\end{abstract}

\section{Introduction}\label{intro}

Tangled magnetic fields are ubiquitous in astrophysical systems, expected to develop in stellar interiors, accretion discs, galaxies and clusters of galaxies \citep{Zeldovich83}. Their ubiquity results from the freezing of magnetic flux into fluid motions in ideal magnetohydrodynamics (MHD): initially straight field lines are quickly tangled by the random stretching motions of a turbulent flow, typically at spatial scales smaller than those associated with the global flow \citep[see, e.g.,][]{Rincon19}. 

If such turbulence is driven sporadically and the subsequent relaxation is viscously dominated (as will be the case when the magnetic Prandtl number $\mathrm{Pm}=\nu/\eta\gg1$), or if the system relaxes to some quasi-stable steady state, then there may be periods of time when dynamical-strength magnetic field tangled at small scales threads relatively quiescent fluid. The large-scale dynamic properties of such a fluid may be highly relevant for questions of how energy is propagated, stored, and dissipated in the astrophysical systems described above. Such considerations motivate an idealised plasma physics problem: how does a static fluid with a statistically homogenous and isotropic Maxwell stress respond to an imposed large-scale impulse? It is this question that we aim to address in this paper, subject to some important simplifying assumptions.

We will treat the problem in the setting of non-resistive MHD, i.e., assuming perfect flux freezing, and thus the invariance of the magnetic field topology. Physically, this means assuming that the small-scale magnetic tangle is itself at scales sufficiently large for magnetic dissipation to be negligible. In pursuit of maximal simplicity, we will also assume that the tangled magnetic field represents an equilibrium state. While this assumption may be unrealistic for MHD turbulence, the dynamics of the tangled equilibrium field are sufficiently complex to warrant investigation, and we anticipate that the lessons learned here will inform study of the more realistic, non-equilibrium case, which we shall report in a later publication.

The plan of this paper is as follows. In Section \ref{homogeneous_tangle}, we formally motivate the problem from the equations of ideal MHD, and describe previous work that modelled the equilibrium Maxwell stress as ``perfectly'' homogeneous and isotropic at all scales. Using this model, \cite{Moffatt86} argued that tangled fields should support waves whose restoring force is the isotropic elasticity afforded by the magnetic field-- these ``magnetoelastic'' waves are the isotropic analogue of Alfv\'{e}n waves on a straight magnetic field. Later, magnetoelastic waves were rederived by \cite{Gruzinov96} in the same approximation, and were suggested (in a somewhat modified form) by \cite{Schekochihin02} and \cite{Maron04} as a possible mechanism for the saturation of the turbulent MHD dynamo. \cite{Williams04} has speculated on the importance of magnetoelasticity in astrophysical systems in which the presence of magnetic fields is traditionally modelled by magnetic viscosity. More recently, \cite{Chen20} have developed a theory of potential vorticity mixing in the solar tachocline, accounting for the effect of a tangled magnetic field elasticity using a similar model. Experimentally, elastic waves have been observed in viscoelastic flows of polymer solutions \citep{Qin19}, which obey a system of equations closely related to MHD \citep{Ogilvie03}.

However, treating the field as perfectly homogeneous naturally precludes the possibility of the large-scale waves driving small-scale motions. In Section \ref{non_homogeneous_tangle}, we show that such motions feed back on the large scales, thereby modifying the dispersion relation of magnetoelastic waves, even when the small scale motions are strongly damped by viscosity (although not, as we will discover, when the small-scale motions are \emph{hyper}viscously damped). In particular, we show that if the field configuration is stable (though this may be an idealisation), this effect always results in a decrease of the wave frequency from the perfectly homogeneous value. 

In Section \ref{sec_fosa}, we develop a mean-field formalism based on an approximation that assumes the coupling of different Fourier modes of the small-scale motions to each other to be small. This treatment is equivalent to the First-Order Smoothing Approximation (FOSA) commonly employed in large-scale dynamo theory (\citealt{Moffatt78, Krause80, Brandenburg05}; see \citealt{Rincon19} for a review), and allows the dispersion relation for magnetoelastic waves to be expressed in terms of statistical properties of the magnetic-field configuration. 
In Section \ref{sec_viscosity}, we use the FOSA to solve the inital-value problem for a magnetoelastic pulse in a viscous fluid. We find that accounting for intermittency gives anomalously fast viscous damping due to the development of small-scale motions.

An important caveat to modelling tangled magnetic fields by equilibrium configurations is that such equilibria appear to be generally \emph{unstable}. Even in the case of periodic, linear force-free equilibria, which have been shown to be stable to wide class of perturbations \citep{Woltjer58a, Molodensky74,Moffatt86}, \cite{East15} have shown that it is often, perhaps always, possible to find ideal perturbations that decrease the total energy. Whether there are any non-trivial cases of periodic magnetostatic equilibria that are stable is an open question. We provide a review of the instability of linear force-free equilibria in Appendix \ref{App:FFinstab}. The theory that we develop here applies to the idealised stable case, or to the case where viscous damping (which proceeds via motions at the scale of the tangled field) is sufficient to delay the onset of the instability.

In Section \ref{numerics}, we present the first (to our knowledge) numerical simulations of isotropic magnetoelasticity. We introduce a sinusoidal velocity perturbation to periodic magnetostatic equilibria, and measure the evolution of the induced standing waves. The results are in excellent agreement with the predictions of our mean-field-theory treatment.



\section{A perfectly homogenous tangle}\label{homogeneous_tangle}

The equations of ideal (non-resistive), incompressible MHD are
\begin{align}
  \displaystyle\frac{\p \boldsymbol{u}}{\p t} + \boldsymbol{u}\bcdot\bnabla\boldsymbol{u}=-\bnabla P& + \boldsymbol{B}\bcdot\bnabla\boldsymbol{B} + \nu\nabla^{2} \boldsymbol{u}, \label{MHDmomentum}\\
  \displaystyle\frac{\p \boldsymbol{B}}{\p t} +\boldsymbol{u}\bcdot\bnabla\boldsymbol{B} &=  \boldsymbol{B}\bcdot\bnabla\boldsymbol{u}, \label{MHDinduction}\\
  \bnabla\bcdot\boldsymbol{u} &= 0, \label{MHDdivu}\\
  \bnabla\bcdot\boldsymbol{B} &= 0, \label{MHDdivB}
\end{align}where $P$ is the total (thermal + magnetic) pressure, determined by \eqref{MHDdivu}; $\nu$ is the kinematic viscosity; $\boldsymbol{u}$ is the fluid velocity; and the magnetic field $\boldsymbol{B}$ is measured in velocity units.

For our purposes, it is convenient to eliminate $\boldsymbol{B}$ from these equations in favour of the Maxwell stress, $M_{ij} = B_i B_j$. Then, in index notation, \eqref{MHDmomentum} becomes
\begin{equation}
  \p_t u_i + u_j\p_j u_i=-\p_i P + \p_j M_{ij}+ \nu \p_j \p_j u_i, \label{Mijmomentum}
\end{equation}while taking the outer product of \eqref{MHDinduction} with $\boldsymbol{B}$ gives
\begin{equation}
  \p_t M_{ij}  = \mathcal{D}_{ij} \left( M , u \right)\equiv M_{ik}\p_k u_j + M_{jk}\p_k u_i - u_k \p_k M_{ij}, \label{Mijinduction}
\end{equation}where $\mathcal{D}_{ij}(M,u)$ is the spatial part of the Lie derivative of $M$ with respect to $\boldsymbol{u}$ -- it is the bilinear operator that gives the rate of change of the tensor $M$ frozen into the flow $\boldsymbol{u}$.

This rewriting of the MHD equations is possible only because we have assumed non-resistive MHD: it would not be possible to write a closed system of evolution equations for $\boldsymbol{u}$ and $M$ if there were a resistive term of the form $\eta \nabla^2 \boldsymbol{B}$ in \eqref{MHDinduction}. Physically, this is because non-resistive MHD is insensitive to the directed nature of magnetic field lines. This is manifest in the fact the Maxwell stress $B_i B_j $ is unchanged by a reversal of the sign of $\boldsymbol{B}$. Magnetic diffusion is, of course, sensitive to such direction reversals; a magnetic-field configuration with a sudden direction reversal can be subject to resistive instabilities.

A simple model of a tangled equilibrium state is obtained by linearising \eqref{Mijmomentum} and \eqref{Mijinduction} about the equilibrium state $M_{ij}= \vM^2\delta_{ij}$, 
where $\vM$ is a constant. A small amount of algebra yields
\begin{equation}
     \frac{\p^2 \boldsymbol{\xi}}{\p t^2} =
  \vM^2 \nabla^2 \boldsymbol{\xi}
  + \nu \nabla^2 \frac{\p \boldsymbol{\xi}}{\p t}, \label{hom_WE}
\end{equation}where $\boldsymbol{\xi}$ is the displacement field defined by $\p_t \boldsymbol{\xi} = \boldsymbol{u}$. This is a wave equation for (viscously damped) magnetoelastic waves with wave speed $\vM$, i.e., waves whose restoring force is the isotropic elasticity of the tangled magnetic field. The corresponding dispersion relation is
\begin{equation}
    \omega = \pm k \vM \sqrt{1+\left(\frac{\nu k}{2 \vM}\right)^2} - \frac{1}{2}i\nu k^2. \label{hom_disprel}
\end{equation}

In the absence of viscosity, magnetoelastic waves have dispersion relation $\omega = \pm k \vM$, and can be thought of as the isotropic equivalent of Alfv\'{e}n waves. Like Alfv\'{e}n waves, magnetoelastic waves are transverse, as $\boldsymbol{\nabla}\bcdot \boldsymbol{u} = 0$ implies $\boldsymbol{k}\bcdot\boldsymbol{\xi}=0$.

The dispersion relation \eqref{hom_disprel} was first obtained via a similar derivation by \cite{Moffatt86}, and later by \cite{Gruzinov96}. However, it is an idealisation because no vector field can satisfy $B_i B_j \propto \delta_{ij}$ at all scales. In the next section, we develop a theory of magnetoelastic waves for a magnetic tangle that is homogeneous and isotropic at large scales, but accounting for its inhomogeneous small-scale structure, and, therefore, the possibility of generating motions at the scale of the magnetic tangle. We find that the dispersion relation \eqref{hom_disprel} is modified by an order-unity factor, even in the limit that the magnetic tangle scale is vanishingly small compared to the scale of the magnetoelastic wave.

\section{Analytic theory of an inhomogenous tangle}\label{non_homogeneous_tangle}

Returning to the MHD equations, we now separate all quantities into large- and small-scale parts, assuming a scale separation between the typical scales of the wave motions and the magnetic tangle. We denote the wavenumbers associated with these scales by $\kw$ and $\kt$, respectively, and take $\kw / \kt \equiv \epsilon $ to be a small parameter. We use the notation $X = \overline{X} + \widetilde{X}$,  where $\overline{X}$ is the spatial average of the quantity $X$ over some intermediate scale that is large compared to the scale of the tangle but small compared to the scale of the wave motion, and $\widetilde{X}$ is the remaining small-scale part.

Denoting equilibrium fields by a subscript zero, we linearise \eqref{Mijmomentum} and \eqref{Mijinduction} about a static equilibrium that satisfies
\begin{equation}
    \overline{M}_{0ij}\equiv \vM^2 \delta_{ij} \label{vE}
\end{equation}for constant $\vM$, while for the moment remaining agnostic about the form of $\widetilde{M}_{0ij}$, but expecting that $\widetilde{M}_0 \sim \overline{M}_0 \sim \vM^2$ because locally there may be an order-unity deviation of the Maxwell stress from its large-scale average. According to this definition, we have $\vM^2 = \frac{1}{3}\overline{v_A^2}$, where $v_A=v_A(\boldsymbol{r})$ is the local Alfv\'{e}n speed, i.e. the speed at which small-scale Alfv\'{e}n waves would propagate along the local magnetic field.

Equation \eqref{vE} implies that the equilibrium Maxwell stress has no structure on the scale of wave motions, therefore our treatment precludes the possibility of treating stochastic fields with structure on all length scales.  We note that, while it is possible to generate a wide class of synthetic stochastic magnetic fields satisfying \eqref{vE}, this assumption may prove too restrictive to model the fields generated by isotropic MHD turbulence. We shall address this and other differences with the turbulent case in a future publication.

The linearised equations are
\begin{equation}
  \p_t^2 \overline{\xi}_i = 
  -\p_i \overline{P} 
  + \vM^2 \p_j \p_j \overline{\xi}_i
  + \p_j \overline{\mathcal{D}}_{ij} ( \widetilde{M}_0 , \widetilde{\xi} )
  + \nu \p_j \p_j \p_t \overline{\xi}_i, 
  \label{xi_bar}
\end{equation}
\begin{equation}
  \p_t^2 \widetilde{\xi}_i = 
  - \p_i \widetilde{P} 
  + \vM^2 \p_j \p_j \widetilde{\xi}_i
  + \p_j \mathcal{D}_{ij} ( \widetilde{M}_0 , \overline{\xi} )
  + \p_j \widetilde{\mathcal{D}}_{ij} ( \widetilde{M}_0 , \widetilde{\xi} )
  + \nu \p_j \p_j \p_t \widetilde{\xi}_i.
  \label{xi_twiddle}
\end{equation}

\subsection{The coupling to small scales is always formally non-negligible \label{neglectSS}}

Let us first ask whether it is possible to find a regime in which the coupling to small scales caused by the $\p_j \overline{\mathcal{D}}_{ij} ( \widetilde{M}_0 , \widetilde{\xi} )$ term in $\eqref{xi_bar}$ can be neglected in comparison with the isotropic restoring force $\vM^2 \p_j \p_j \overline{\xi}_i$. This would require $\kt\,\widetilde{\xi} \ll \kw\,\overline{\xi}$, i.e., $\widetilde{\xi} \ll \epsilon \, \overline{\xi}$. Under this ordering, there are no terms in \eqref{xi_twiddle} that can balance $\p_j \mathcal{D}_{ij} ( \widetilde{M}_0 , \overline{\xi} )$ apart from the viscous term. Such a balance implies $\kw \kt \vM^2 \overline{\xi} \sim \nu \kt^2 \,\omega \widetilde{\xi} \ll \nu \kt^2 \,\omega \epsilon\, \overline{\xi}$, i.e., $\vM^2 \ll \nu \omega$. Assuming that the large-scale response is indeed elastic gives $\omega \sim k \vM$ as before, which leaves us with $\vM \ll \nu \kw$. However, this is precisely the condition for the magnetoelastic wave's viscous damping rate to be large compared to the wave frequency, a contradiction to the assumed scaling $\omega \sim k \vM$. Hence we find that \emph{the coupling to small scales is always non-negligible if the field is to respond elastically, i.e., if} $\omega \sim k \vM$.

This conclusion is a result of the arithmetics of powers of $k$ in each of the terms in \eqref{xi_bar} and \eqref{xi_twiddle}: essentially, viscosity with its $k^2$ scaling does not `switch on' fast enough at larger $k$ to prevent the driving of dynamically important small-scale motions. Quenching them requires a viscous damping that scales with $k$ \emph{faster} than $k^2$, i.e., a hyperviscosity. In Section \ref{sec:hyperviscous}, we will present numerical experiments with tangle scales hyperviscously damped, which do indeed show precise agreement with  \eqref{hom_disprel}.

That the coupling to small scales should be formally non-negligible is, in fact, clear on intuitive physical grounds; a set of disconnected `blobs' of magnetic field may well satisfy $\overline{M}_{0ij} \propto \delta_{ij}$, but will not be able to support a net tension on scales much larger than the typical blob size. In this case, the term describing the coupling to small scales $\p_j \overline{\mathcal{D}}_{ij} ( \widetilde{M}_0 , \widetilde{\xi} )$ in $\eqref{xi_bar}$ will be non-negligible and its effect will be to cancel the large-scale elasticity term, $\vM^2 \p_j \p_j \overline{\xi}_i$. This intuition also suggests that the effect of the coupling to small scales should \emph{reduce} the effective elasticity, an expectation that is confirmed by the analysis in the next section. Hyperviscosity modifies this picture by preventing any differential motion on small scales, so the fluid behaves as though the magnetic blobs were connected by rigid rods that allow the large-scale tension to be maintained.

\subsection{Normal-mode analysis \label{sec:normalmodes}}

Some general statements can be made regarding the elastic response of an inviscid tangle by conducting a normal-mode analysis. The essential result of this section is that for a stable field configuration, the frequency of magnetoelastic waves is always decreased from $\kw \vM$ as a result of intermittency of the Maxwell stress.

We expand the displacement field and Maxwell stress in Fourier modes, viz., $\xi_i=\sum_{\boldsymbol{k}} \xi_i(\boldsymbol{k})e^{i\left(\boldsymbol{k}\bcdot\boldsymbol{r}-\omega t\right)}$ and $M_{0ij}=\sum_{\boldsymbol{k}} M_{0ij}(\boldsymbol{k})e^{i\boldsymbol{k}\bcdot\boldsymbol{r}}$, define the projection operator $\mathcal{P}_{ij}(\bk)=\delta_{ij}-k_i k_j/k^2$, and define the matrix elements $A_{ij}(\bk, \bkp)$ of the operator $\mathcal{P}_{il}\p_j \mathcal{D}_{lj} \left( M_0 , \,\bullet\,\right)$ in the Fourier basis so that
\begin{equation}
  -\omega^2 \xi_i(\bk) = 
  \sum_{\bkp} A_{ij}(\bk, \bkp) \xi_j(\bkp).
  \label{normalmodesystem}
\end{equation}
The explicit form of $A_{ij}(\bk, \bkp)$ can be obtained straightforwardly from the definition \eqref{Mijinduction} of $\mathcal{D}$, viz.,
\begin{align}
    A_{in}(\bk, \bkp) &= - \mathcal{P}_{il}\left(\boldsymbol{k}\right)k_j \left[-\left(k_m-k'_m\right) M_{0lj}\left(\boldsymbol{k}-\boldsymbol{k}'\right)\mathcal{P}_{mn}\left(\bkp\right) \right.\nonumber\\
    &\quad + \left. M_{0lm}\left(\boldsymbol{k}-\boldsymbol{k}'\right)k_m' \mathcal{P}_{jn}\left(\bkp\right)+M_{0jm}\left(\boldsymbol{k}-\boldsymbol{k}'\right)k_m' \mathcal{P}_{ln}\left(\bkp\right)\right].
  \label{fourier}
\end{align} Importantly, $A_{ij}(\bk, \bkp)$ is Hermitian, i.e., $A_{ij}(\bk, \bkp)=\left[ A_{ji}(\bkp, \bk) \right]^{*}$\footnote{This is a consequence of a general result in MHD that the linearised force operator $\boldsymbol{F}$, defined by $\rho_0\, \p_t^2 \boldsymbol{\xi}=\boldsymbol{F}\left[\boldsymbol{\xi}\right]$, is self-adjoint \citep[see][]{Kulsrud05}, viz., for any $\boldsymbol{\xi}$ and $\boldsymbol{\eta}$, 
$\int\text{d}^3 \boldsymbol{r}\, \boldsymbol{\eta}\cdot \boldsymbol{F}\left[\boldsymbol{\xi}\right] = \int\text{d}^3 \boldsymbol{r}\, \boldsymbol{\xi}\cdot \boldsymbol{F}\left[\boldsymbol{\eta}\right]$.}. Let us prove this explicitly. The Hermitian conjugate of \eqref{fourier} is
\begin{align}
    \left[ A_{ni}(\bkp, \bk) \right]^{*} &= - \mathcal{P}_{nl}\left(\bkp\right)k_j' \left[-\left(k'_m-k_m\right) M_{0lj}\left(\boldsymbol{k}-\boldsymbol{k}'\right)\mathcal{P}_{mi}\left(\bk\right) \right.\nonumber\\
    &\quad + \left. k_m M_{0lm}\left(\boldsymbol{k}-\boldsymbol{k}'\right) \mathcal{P}_{ji}\left(\bk\right)+k_m M_{0jm}\left(\boldsymbol{k}-\boldsymbol{k}'\right) \mathcal{P}_{li}\left(\bk\right)\right].
  \label{Ani}
\end{align} Taking $m\leftrightarrow l$ in the first term of \eqref{fourier}, and $j\leftrightarrow l$ in the second term, then subtracting \eqref{Ani} from \eqref{fourier} gives
\begin{align}
    A_{in}(\bk, \bkp)&-\left[ A_{ni}(\bkp, \bk) \right]^{*}  = \nonumber \\ & \mathcal{P}_{il}\left(\boldsymbol{k}\right) \mathcal{P}_{nj}\left(\bkp\right)\left(\delta_{lp}\delta_{jq}-\delta_{lq}\delta_{jp}\right)
    \left(k_q-k'_q\right)\left(k_m-k'_m\right) M_{0pm}\left(\boldsymbol{k}-\boldsymbol{k}'\right),
\end{align} which is zero, as $\delta_{lp}\delta_{jq}-\delta_{lq}\delta_{jp}$ = $\epsilon_{rlj}\epsilon_{rpq}$, and $\epsilon_{rpq} \left(k_q-k'_q\right)\left(k_m-k'_m\right) M_{0pm}\left(\boldsymbol{k}-\boldsymbol{k}'\right) = 0$ because the equilibrium field must satisfy $\boldsymbol{\nabla}\times \left(\boldsymbol{\nabla}\bcdot\boldsymbol{M}_0\right)=0$ (see equation \ref{Mijmomentum}). Therefore, $A_{ij}(\bk, \bkp)=\left[ A_{ji}(\bkp, \bk) \right]^{*}$, q.e.d.

When $\boldsymbol{k}'=\boldsymbol{k}$, \eqref{fourier} reduces to $A_{ij}(\bk,  \bk)=-M_{0ll}(\boldsymbol{0})k^2 \delta_{ij}=-k^2\vM^2 \delta_{ij}$, so each Fourier mode is subject to the large-scale isotropic restoring force, as in \eqref{xi_twiddle}. Terms with $\boldsymbol{k}'\neq\boldsymbol{k}$ describe the coupling of different Fourier modes: indeed, from \eqref{fourier}, we see that Fourier modes of the displacement field $\boldsymbol{\xi}$ with wavevectors $\boldsymbol{k}$ and $\boldsymbol{k}'$ are coupled to each other only when $M_{0jm}\left(\boldsymbol{k}-\boldsymbol{k}'\right)$ is non-zero. This observation shows that two large-scale Fourier modes with $\boldsymbol{k},\,\boldsymbol{k}'\sim\kw$ are not coupled by \eqref{fourier}, because $M_{0ij}(\bk)=0$ for $\boldsymbol{k}\sim\kw$, by \eqref{vE}. In principle, these large-scale Fourier modes can still be coupled as a result of each of them coupling individually to small scales. The condition for two large-scale Fourier modes to be coupled in this way is the existence of a path in Fourier-space between them, along the wavevectors of $M_0$. This is precisely the condition for the Fourier modes of $M_0$ to `beat' at the magnetoelastic-wave scale, which is in contradiction to the assumption of statistical homogeneity at large scales. Henceforth, we will assume that such beating is absent from $M_0$\footnote{Equivalently, we assume that all positive integer powers of $M_0$ have no large-scale structure.}, and, therefore, that different large-scale Fourier modes are completely decoupled. We will also assume that large-scale modes with the same $\boldsymbol{k}$ but different spatial directions are decoupled, which is a natural consequence of statistical isotropy. This discussion implies that it is sensible to decompose the large-scale perturbation into its constituent Fourier modes, each of which will independently drive small-scale motions that feed back on the particular large-scale Fourier mode that caused them, but not on any other large-scale modes.

If we take $M_0$ to have a finite number of non-zero Fourier modes\footnote{Any field configuration can be approximated to arbitrary accuracy by making this number large.},
then $A_{ij}(\bk, \bkp)$ can be considered as a Hermitian \emph{matrix} whose elements  $A_{(\bk, i)(\bkp,\, j)}$ describe the coupling of the mode $\xi_i \left(\bk\right)$ with $\xi_j \left(\bkp\right)$. Since a large-scale mode may couple to small-scale modes but \emph{not} to other large-scale modes, the matrix representation of $A$ is block diagonal, with each block corresponding to one particular large-scale mode and the small-scale modes to which it couples. Let $\mathcal{A}$ be the block of size  $(N+1)\times(N+1)$ corresponding to a particular large-scale mode $\xi_z (\bkw)$, taken to be in the $z$-direction without loss of generality, that is coupled to $N$ small-scale modes. The general structure of $\mathcal{A}$ is
\begin{equation}
    \mathcal{A} = 
\begin{pmatrix}
-\kw^2 \vM^2 &  \boldsymbol{f}^\dag\\
 \boldsymbol{f} & \mathcal{B}
\end{pmatrix}.
\end{equation} where the element $\mathcal{A}_{(\bkw, z)(\bkw, z)}\equiv A_{zz}(\bkw, \bkw)=- \kw^2 \vM^2$ describes the isotropic elastic restoring force on the large-scale mode; $\mathcal{A}_{(\bk, i)(\bkw, z)}\equiv A_{iz}(\bk, \bkw)\equiv f_{(\bk, i)}$ for $|\bk|\sim \kt$ is an $N$-dimensional vector that gives the coupling of the small-scale modes to the large-scale mode; and $\mathcal{A}_{(\bk, i)(\bkp,\, j)}\equiv A_{ij}(\bk,\bkp)\equiv \mathcal{B}_{(\bk, i)(\bkp,\, j)}$ for $|\bk|,\,|\bkp|\sim\kt$ is the $N\times N$ Hermitian matrix representing the coupling of the relevant small-scale modes to each other. The sizes of these three components are respectively $\kw^2 \vM^2$, $\kw \kt \vM^2$ and $\kt^2 \vM^2$.

The $N+1$ normal modes of the system have frequencies $\omega_\mu$ satisfying $\det \left(\mathcal{A}+\omega_{\mu}^2 I_{N+1}\right)=0$, where $I_{N+1}$ is the identity matrix of size $N+1$. To leading order in $\epsilon$, the fast, small-scale motions with frequency $\omega_\mu\sim \kt \vM $ satisfy
\begin{equation}
    \det 
\begin{pmatrix}
\omega_{\mu}^2 &  \boldsymbol{f}^\dag\\
 \boldsymbol{f} & \mathcal{B} + \omega_{\mu}^2 I_N
\end{pmatrix}=
\omega_{\mu}^2 \det
\left(\mathcal{B}+\omega_{\mu}^2 I_N\right)=0\implies \det \left(\mathcal{B}+\omega_{\mu}^2 I_N\right)=0,\label{eqnforomegafast}
\end{equation} so that the fast frequencies are unaffected by the coupling of large and small scales. Equation \eqref{eqnforomegafast} has $N$ solutions for $\omega_{\mu}^2$, which we denote by $\mu=1,...,N$.
The remaining slow, large-scale solution with frequency $\omega_0\sim\kw \vM$ represents a magnetoelastic wave, and satisfies
\begin{equation}
    \det 
\begin{pmatrix}
\omega_0^2- \kw^2 \vM^2 &  \boldsymbol{f}^\dag\\
 \boldsymbol{f} & \mathcal{B}
\end{pmatrix}=0.\label{eqnforomega}
\end{equation}
Defining the block matrix
\begin{equation}
    T=\begin{pmatrix}
  1 & 0\\
  0 & U
\end{pmatrix},
\end{equation} where $U$ is the $N\times N$ unitary matrix that diagonalises the matrix $\mathcal{B}$, and using the invariance of the determinant under the basis transformation defined by $T$, we obtain
\begin{equation}
    \det 
\begin{pmatrix}
\omega_0^2- \kw^2 \vM^2 &  \left(U\boldsymbol{f}\right)^\dag\\
\, U\boldsymbol{f} & \Lambda
\end{pmatrix}=0,\label{detmat=0}
\end{equation} where $\Lambda=U\mathcal{B}\,U^\dag=\text{diag}\left(\lambda_1, \,\dots,\lambda_N \right)=\text{diag}\left(-\omega_1^2, \,\dots,-\omega_N^2 \right)$ is the diagonal matrix of eigenvalues of $\mathcal{B}$, which are the $N$ (negative squared) frequencies of the small-scale system. Since $\Lambda$ is diagonal, the determinant is simple to evaluate, and \eqref{detmat=0} leads to
\begin{equation}
    \omega_0^2 = \kw^2 \vM^2 - \sum_{\mu=1}^N \frac{\large|\left(U \boldsymbol{f}\right)_\mu\large|^2}{\omega_\mu^2}.\label{fullomegasystem}
\end{equation}Equation \eqref{fullomegasystem} shows that the effect of the coupling to small-scale motions is to \emph{reduce} the frequency of large-scale waves, as long as the equilibrium is stable, i.e., as long as all $\omega_{\mu}^2>0$. Furthermore, since the components of $U$ and $\boldsymbol{f}$ are $\sim 1$ and $\sim \kw \kt \vM^2$ respectively, while $\omega_{\mu}\sim \kt \vM$ for $\mu \geq 1$, the frequency is reduced by a factor of order unity. Physically, this is akin to the elastic response of a tangled ball of elastic string compared to a solid elastic block: a deformation will generally produce a smaller restoring force in the elastic ball because the strings can move relative to each other to reduce the elastic energy (this is not possible in a hyperviscous fluid as small-scale motions are suppressed).

Of course, there is no reason that the right-hand side of \eqref{fullomegasystem} should be positive; if it is not, \eqref{fullomegasystem} describes a growing perturbation at large scales. This result shows that it is in principle possible for an \emph{unstable} magnetic tangle to relax via motions on \emph{large} scales rather than small scales. However, we have found no example of this in our numerical studies, where the effect of the tangled field on large-scale modes was always restoring, with instability proceeding only via motions at small scales.

The eigenvectors of $\mathcal{A}$ can be obtained to leading order by noting that $\mathcal{A} \left(0, \boldsymbol{e}_{\mu}\right)^\text{T} = \lambda_\mu\left(0, \boldsymbol{e}_{\mu}\right)^\text{T} + O\left(\epsilon\right)$, where $\boldsymbol{e}_{\mu}$ is the $\mu$th eigenvector of the matrix $\mathcal{B}$. The final eigenvector with associated frequency $\omega_0$ is uniquely constrained by orthogonality to be $\left(1, \boldsymbol{0}\right)^\text{T}+O\left(\epsilon\right)$. Therefore, a magnetoelastic wave consists primarily of a large scale oscillation, together with small-amplitude (vanishing as $\epsilon\rightarrow0$), slow (frequency $\omega_0$), small-scale oscillations. Physically, these small-scale oscillations represent the rearrangement of small-scale structures to reduce the elastic energy and, therefore, the large-scale tension. Despite having small amplitude, these small-scale motions are by no means negligible -- as we have found, the magnetoelastic wave frequency is changed by a factor of order unity in their absence.


\subsection{The First-Order Smoothing Approximation (FOSA) \label{sec_fosa}}

In this section, we describe an approximate method to obtain the wave frequency $\omega_0$ in terms of statistical properties of the magnetic tangle, by assuming that the coupling between different small-scale Fourier modes is small. This approximation is equivalent to neglecting the term $\p_j \widetilde{\mathcal{D}}_{ij} ( \widetilde{M}_0 , \widetilde{\xi})$ in \eqref{xi_twiddle}; in the context of large-scale kinematic dynamo theory, it is often called the First-Order Smoothing Approximation (FOSA) or the Second-Order Correlation Approximation (SOCA). In the dynamo-theory context, it is employed in the small-scale part of the induction equation to neglect a similar `fluctuating part of the product of two fluctuations' term, allowing the small-scale induction equation to be solved for the small-scale magnetic field, which can then be used to compute the growth of the large-scale field \citep[for a review, see][]{Rincon19}. Our use of this approximation is directly complementary: instead of solving an equation for the small-scale magnetic field given a prescribed flow, we use the FOSA to solve an equation for a small-scale flow given a known small-scale magnetic-field configuration, $M_0$.

Much like in the dynamo-theory context, where the FOSA is rigorously justified only when either the velocity correlation time is small, or $\mathrm{Rm}\ll1$, the assumption of weak coupling between modes is unlikely to be well satisfied in any real magnetic tangle. However, the FOSA remains a useful tool, and we show in Section \ref{numerics} that it provides a remarkably good description of numerical simulations of large-scale waves in tangled-magnetic-field equilibria.

Under the FOSA, we neglect all off-diagonal terms of the matrix $\mathcal{B}$ in \eqref{eqnforomega}, so that the coupling of large and small scales is retained, but the small-scale modes do not couple to each other. This simplifies \eqref{fullomegasystem} considerably because now $U=I_N$ and $ \omega_\mu= k_\mu \vM $ for $\mu=1,...,N$, where $k_\mu$ are the wavenumbers of the $N$ small-scale modes, so that \eqref{fullomegasystem} becomes
\begin{equation}
    \omega_0^2 = k^2 \vM^2 - \sum_{\bkp\neq\bkw} \frac{A_{zp}(\bkw, \bkp)A_{pz}(\bkp, \bkw) }{k'^2 \vM^2}\label{FOSAo0}.
\end{equation}

With $i \rightarrow z$ and $\bk\rightarrow \bkw$, \eqref{fourier} gives
\begin{align}
    A_{zn}(\bkw, \bkp) &= - \mathcal{P}_{zl}\left(\bkw\right)k_{\text{w}j} \left[-\left(k_{\text{w}m}-k'_{m}\right) M_{0lj}\left(\bkw-\boldsymbol{k}'\right)\mathcal{P}_{mn}\left(\bkp\right) \right.\nonumber\\
    &\quad + \left. M_{0lm}\left(\bkw-\boldsymbol{k}'\right)k_m' \mathcal{P}_{jn}\left(\bkp\right)+M_{0jm}\left(\bkw-\boldsymbol{k}'\right)k_m' \mathcal{P}_{ln}\left(\bkp\right)\right].
\end{align}With $|\bkp| \sim \kt$, the first term in the square brackets vanishes to leading order in $\epsilon$ because $k'_m \mathcal{P}_{mn}\left(\bkp\right)=0$. Relabelling $n\rightarrow p$, then swapping $m\leftrightarrow j$ and $m\leftrightarrow l$ in the second term, we obtain 
\begin{align}
    A_{zp}(\bkw, \bkp) &= - \mathcal{P}_{zl}\left(\bkw\right)k_{\text{w}m} M_{0lj}\left(\bkw-\boldsymbol{k}'\right)k_j' \mathcal{P}_{mp}\left(\bkp\right)\nonumber\\&\quad-\mathcal{P}_{zm}\left(\bkw\right)k_{\text{w}j} M_{0jl}\left(\bkw-\boldsymbol{k}'\right)k_l' \mathcal{P}_{mp}\left(\bkp\right)',\nonumber \\
  &= - k_{\text{w}n} k_l' \mathcal{P}_{zq}\left(\bkw\right)\mathcal{P}_{mp}\left(\bkp\right)\left(\delta_{nm}\delta_{jq}+\delta_{nj}\delta_{mq}\right)M_{0lj}\left(\bkw-\boldsymbol{k}'\right).\label{Azp}
\end{align}An analogous expression for $A_{pz}(\bkp, \bkw)$ can be obtained directly from \eqref{Ani}, but the derivation is much more involved. Instead, we can use the fact that $A$ is Hermitian:
\begin{align}
    A_{pz}(\bkp, \bkw)&=\left[A_{zp}(\bkw, \bkp)\right]^{*}\nonumber\\&=- k_{\text{w}n} k_l' \mathcal{P}_{zq}\left(\bkw\right)\mathcal{P}_{mp}\left(\bkp\right)\left(\delta_{nm}\delta_{jq}+\delta_{nj}\delta_{mq}\right)M_{0lj}\left(\boldsymbol{k}'-\bkw\right)\nonumber \\
    &=- k_{\text{w}n} k_j' \mathcal{P}_{zm}\left(\bkw\right)\mathcal{P}_{qp}\left(\bkp\right)\left(\delta_{nq}\delta_{lm}+\delta_{nl}\delta_{mq}\right)M_{0lj}\left(\boldsymbol{k}'-\bkw\right),\label{Apz}
\end{align}where to obtain the final expression we have swapped $l \leftrightarrow j$ and $q \leftrightarrow m$.
With \eqref{Azp} and \eqref{Apz}, \eqref{FOSAo0} becomes:
\begin{align}
    \omega_0^2 = \kw^2 \vM^2 - \left(\delta_{nm}\delta_{jz}+\delta_{nj}\delta_{mz}\right)\left(\delta_{zl'}\delta_{n'q'}+\delta_{l'n'}\delta_{zq'}\right)k_{\text{w}n} k_{\text{w}n'}R_{mq'jl'},\label{deltas}
\end{align}where 
\begin{equation}
    R_{mq'jl'}=\sum_{\bkp} \frac{k_l' k_{j'}'}{k'^2 \vM^2} \mathcal{P}_{q'm}\left(\bkp\right)\widetilde{M}_{lj}\left(-\bkp\right)  \widetilde{M}_{l'j'}\left(\bkp\right), \label{R}
\end{equation} to leading order in $\epsilon$. The rank-four tensor $R_{pq'jl'}$ is a statistical property (in the sense of volume averaging) of the magnetic tangle. Since we have assumed statistical isotropy, the most general form it can take is $R_{pq'jl'}=a\delta_{pq'}\delta_{jl'}\,+\,b\delta_{pj}\delta_{q' l'}\,+\,c\delta_{pl'}\delta_{q'j}$. Noting from \eqref{R} that $R_{pq'jl'}$ is symmetric in $p$ and $q'$ and vanishes on contraction of $p$ with $l'$, we find
\begin{equation}
    R_{pq'jl'}=\frac{1}{30\vM^4}\left(4\delta_{l'j}\delta_{pq'}-\delta_{l'p}\delta_{jq'}-\delta_{l'q'}\delta_{jp}\right)\left(\left\langle P_0^2\right\rangle - \left\langle P_0\right\rangle^2\right).
\end{equation}where we have used the equilibrium condition $k_j M_{0ij}\left(\bk\right) + k_i P_0\left(\bk\right)=0$, where $P_0$ is the equilibrium total pressure distribution, and identified
\begin{equation}
    \sum_{\bkp}\left|\widetilde{P}_0 \left(\bkp\right)\right|^{2} = \left(\left\langle P_0^2\right\rangle - \left\langle P_0\right\rangle^2\right), 
\end{equation}where angled brackets indicate spatial averages.
On substitution of this result into \eqref{deltas} and contraction of the many Kronecker deltas, we finally arrive at
\begin{equation}
\omega_0^2 = k^2 \vM^2\left(1-\chi\right),
\end{equation}where
\begin{equation}
    \chi = \frac{1}{5\vM^4}\left(\left\langle P_0^2\right\rangle - \left\langle P_0\right\rangle^2\right). \label{chi}
\end{equation}We therefore find that under the FOSA, the effect of the small-scale structure of the magnetic tangle is to reduce the frequency of magnetoelastic waves by an amount proportional to the variance of the total pressure. This is really a statement about the magnetic field, since the total pressure must balance the magnetic tension force in equilibrium, so \eqref{chi} shows that \emph{more intermittent magnetic fields are less elastic}. This is precisely in agreement with the intuitive reasoning of Section \ref{neglectSS}, where we argued that a field configuration consisting of disconnected magnetic `blobs' would not support a large-scale tension -- such a field would have a large value of $\chi$.

In the special case of a force-free magnetic tangle, i.e., when only magnetic pressure balances the magnetic tension, $P_0=B_0^2/2$, so
\begin{equation}
    \chi = \frac{9}{20}\left(\frac{\left\langle B_0^4\right\rangle}{\left\langle B_0^2\right\rangle^2} - 1\right).\label{chiFOSA}
\end{equation}


\subsection{Magnetoelastic waves in a viscous fluid \label{sec_viscosity}}

In this section, we develop a theory for magnetoelastic waves propagating through a viscous fluid. As we found in Section \ref{sec:normalmodes}, a magnetoelastic wave in an inviscid fluid is a large-scale oscillation, accompanied by slow ($\omega\sim \kw\vM$), small-amplitude oscillations associated with the relaxation of small-scale structures in response to the large-scale tension. In the viscous case, these small-scale structures will be damped by viscosity; their presence therefore results in an anomalous viscous damping of magnetoelastic waves.

For definiteness, we take the initial condition to be a large-scale velocity perturbation $u_z$ along $z$ with wavevector $\bkw$. We solve the inital value problem by taking a Laplace transform in time:
\begin{equation}
  p^2 \xi_z(\bkw) - u_z\left(\bkw,t=0\right) = -\kw^2\vM^2 \xi_z(\bkw)+
  \sum_{\bkp} A_{zj}(\bkw, \bkp) \xi_j(\bkp)
  +\nu \kw^2 p \xi_z(\bkw). \label{viscous_startpoint}
\end{equation} where $p$ is the Laplace conjugate variable to time (analogous to $i\omega$)

In the FOSA, an inititally unperturbed small-scale mode with wavenumber $ k' \gg \kw$ satisfies 
\begin{equation}
  p^2 \xi_j(\bkp) = -\vM^2 k'^2 \xi_j(\bkp) +
  A_{jz}(\bkp, \bkw) \xi_z(\bkw)
  +\nu k'^2 p \xi_j(\bkp).
\end{equation}
Solving this algebraic equation for $\xi_j(\bkp)$, and substituting into \eqref{viscous_startpoint}, we obtain
\begin{align}
  p^2 \xi_z(\bkw) - u_z\left(\bkw,t=0\right) &=-\kw^2\vM^2 \xi_z(\bkw)\nonumber\\&\quad+
  \sum_{\bkp}  \frac{A_{zj}(\bkw, \bkp) A_{jz}(\bkp, \bkw)}{p^2 + \vM^2 k'^2 + \nu k'^2 p}\xi_z (\bkw)
  +\nu \kw^2 p \xi_z(\bkw). \label{A10}
\end{align}where the sum is over all small-scale modes. 

With $p\sim\omega_0\sim \kw\vM$, the $p^2$ in the denominator of the coupling term in \eqref{A10} is small compared to $\vM^2 k'^2$\footnote{The $p^2$ is only non-negligible if the large-scale Fourier mode under consideration has motions on the \emph{fast} timescale, i.e., $p\sim \kt \vM$. In the inviscid case, we found that a purely large-scale perturbation is \emph{almost} a normal mode -- the true (slow) normal mode is a large-scale perturbation with accompanying small-amplitude (vanishing as $\epsilon\rightarrow0$) small-scale perturbations. Therefore, a purely large-scale initial perturbation can be decomposed into the slow normal mode and small-amplitude fast modes. Similarly, fast modes are \emph{almost} pure small-scale motions, but do have an accompanying small-amplitude \emph{large}-scale motion. This means that in principle, there will be a fast, large-scale response to an imposed perturbation at large scales, but its amplitude will be vanishingly small compared to the amplitude of the slow response. In the presence of viscosity, we do not expect this conclusion to be modified -- intuitively, viscosity should further suppress the amplitude of fast motions. In Appendix \ref{App:LaplaceAmpl}, we verify this conclusion by explicitly showing that the amplitude of any Laplace mode with $p\sim \kt \vM$ vanishes in the limit $\epsilon \rightarrow 0$.}.
Neglecting it, we can use  \eqref{Azp} and \eqref{Apz} analogously to the inviscid case to write
\begin{equation}
  p^2 \xi_z(\bkw) - u_z\left(\bkw,t=0\right) =-\kw^2\vM^2 \xi_z(\bkw)+\frac{\chi}{1 + \nu p/\vM^2}\xi_z(\bkw)
  +\nu \kw^2 p \xi_z(\bkw).
\end{equation}Note that, as predicted in Section \ref{neglectSS}, the term representing coupling to small scales remains finite as $\epsilon\rightarrow0$, and can only be neglected when $\nu p / \vM^2 \sim \nu \kw^2 / \kw \vM \gg 1$, i.e., when wave motions are strongly damped. In contrast, in the hyperviscous case, the number of powers of $k'$ is increased in the viscosity term in the denominator of the coupling term in \eqref{A10}, so that it is sufficient for \emph{small} scales to be strongly damped for the coupling term to be negligible as $\epsilon\rightarrow0$. We therefore recover the simple wave equation \eqref{hom_WE} in the hyperviscous case, as predicted in Section \ref{neglectSS}.

Finally, we invert the Laplace transform using Cauchy's residue theorem. The solution is a sum of Laplace modes $\overline{u}_i \left(t\right) = \left(\sum_n A_n e^{p_n t}\right) \overline{u}_i \left(0\right)$ with amplitudes
\begin{equation}
 A_{n} = \text{Res} \left[ \frac{p}{D(p
 )} , p\rightarrow p_n \right], \label{residue}
\end{equation}where `$\mathrm{Res}$' denotes the residue and $\{p_n\}$ are the roots of the dispersion relation
\begin{equation}
    D(p_n) \equiv p_n^2+\left(1-\frac{\chi}{1 + \nu p_n/\vM^2}\right) \vM^2k^2+\nu k^2 p_n = 0. \label{viscousdisprel}
\end{equation}It is instructive to consider the case where the viscous damping rate is finite but small compared to the wave frequency, i.e., $\hat{\nu}\equiv \nu \kw/\vM\ll1$. The solutions of \eqref{viscousdisprel} in this limit are
\begin{align}
    p_1 & = +i \vM \kw \sqrt{1-\chi} - \frac{1}{2}\nu \kw^2 \left(1+\chi\right) + \textit{O}\left(\hat{\nu}^2\right), \label{p1}\\
    p_2 & = -i \vM \kw \sqrt{1-\chi} - \frac{1}{2}\nu \kw^2 \left(1+\chi\right) + \textit{O}\left(\hat{\nu}^2\right), \label{p2}\\
    p_3 & = -\vM^2 /\nu + \textit{O}\left(\hat{\nu}\right)\label{p3},
\end{align} with corresponding amplitudes
\begin{equation}
    A_{1}  =  \frac{1}{2} +  \textit{O}\left(\hat{\nu}\right), \quad
    A_{2}  =  \frac{1}{2} +  \textit{O}\left(\hat{\nu}\right), \quad
    A_{3}  = \chi \hat{\nu}^2 + \textit{O}\left(\hat{\nu}^3 \right).
\end{equation}

In the first two solutions, $p_1$ and $p_2$, given by \eqref{p1} and \eqref{p2}, we find the wave modes of the previous section, but now with a damping rate that is larger than in the perfectly homogeneous case by a factor of $1+\chi$. This increased damping rate is expected: it is a consequence of the viscous damping of the small-scale motions associated with magnetoelastic waves. It may appear odd that a damping associated with small-scale motions has a rate $\propto \nu \kw^2$, especially as the condition $\hat{\nu}\ll1$ does not exclude the possibility of fast, small scale motions being over-damped by viscosity. The resolution is that the small-scale motions associated with a magnetoelastic wave are \emph{slow} (frequency $\sim \kw \vM$), not fast, and therefore they are not viscously dominated when $\hat{\nu}\ll1$, as $\nu \p_j \p_j \p_t \widetilde{\xi}_i\sim\nu \kt^2 \omega_0 \widetilde{\xi} \sim\hat{\nu}\kt^2\vM^2 \widetilde{\xi} \ll \vM^2 \p_j \p_j \widetilde{\xi}_i\sim  \kt^2\vM^2 \widetilde{\xi} $. These motions satisfy the balance $\vM^2 \p_j \p_j \widetilde{\xi}_i
  \sim \p_j \mathcal{D}_{ij} ( \widetilde{M}_0 , \overline{\xi} )$, which gives $\widetilde{\xi}\sim\epsilon\overline{\xi}$. The rate of energy dissipation associated with them is therefore $\sim\nu \kt^2 \omega_0^2 \widetilde{\xi}^2\sim\nu \kw^2 \omega_0^2 \overline{\xi}$, which is consistent with the anomalous damping rate in \eqref{p1} and \eqref{p2}.

The third mode, $p_3$, given by \eqref{p3}, describes the case where the slow, small-scale motions associated with the waves \emph{do} become viscously-dominated, so obey the balance $\vM^2 \p_j \p_j \widetilde{\xi}_i
  \sim \kt^2 \p_t \widetilde{\xi}$, giving a damping rate of $\vM^2/\nu$. As above, these motions become viscously-dominated only when $\hat{\nu}\sim 1$, explaining the vanishing amplitude of this mode when $\hat{\nu}\ll1$\footnote{In the FOSA, there is only one mode of this type because all overdamped small-scale motions relax at the same rate, $\vM^2/\nu$. This is because, in the notation of Section \ref{sec:normalmodes}, the equation of motion of the over-damped, small-scale system is $\nu k^2 \p_t \widetilde{\xi}_i(\bk)=\sum_{(\bkp,\, j)}\mathcal{B}_{(\bk, i)(\bkp,\, j)} \widetilde{\xi}_j(\bkp)$. In the FOSA, $\mathcal{B}$ is diagonal, with elements $\mathcal{B}_{(\bk, i)(\bk,\, i)} \propto k^2$, so that the viscous relaxation time has no $k$ dependence.  In the exact system, $\mathcal{B}$ is not diagonal, lifting the degeneracy. Then, $p_3$ is replaced by many Laplace modes for each of the possible viscous relaxation timescales, all with frequencies $\sim \vM^2/\nu$.}.

\section{Numerical study}\label{numerics}

\subsection{Ideal instability of tangled magnetic field equilibria\label{subsec:instabintro}}

The analysis in the preceeding sections has been idealised because real tangled magnetic equilibria are generically \emph{unstable}, even to ideal perturbations \citep{Er-Riani14,East15}. The instability typically proceeds via motions at small scales that are fast compared to the magnetoelastic wave motions, but are inhibited by the presence of viscosity. However, the growth rate associated with viscously-dominated, small-scale unstable modes is $\sim \vM^2/\nu$, which becomes comparable with the wave frequency when $\hat{\nu}=\nu \kw^2 / \kw \vM \sim 1$, i.e., when magnetoelastic waves are strongly damped by viscosity. The question then arises as to whether any static, tangled-magnetic-field equilibrium will pesist long enough for waves to propagate through it, even with strong viscosity. In fact, we have found in our numerical study that when the equilibrium state is a \emph{linear force-free} magnetic field (see next section) the growth rate of the instability is sufficiently slow for the equilibrium configuration to persist for many wave periods.

The existence of an instability of linear force-free magnetic fields in ideal MHD has only recently been appreciated, and indeed many inaccurate statements have historically been presented in the literature. For the interested reader, we present a short review of this instability in Appendix \ref{App:FFinstab}.

\subsection{Simulation setup}\label{simulating}

\begin{figure}
  \centering
  \includegraphics[scale=0.57]{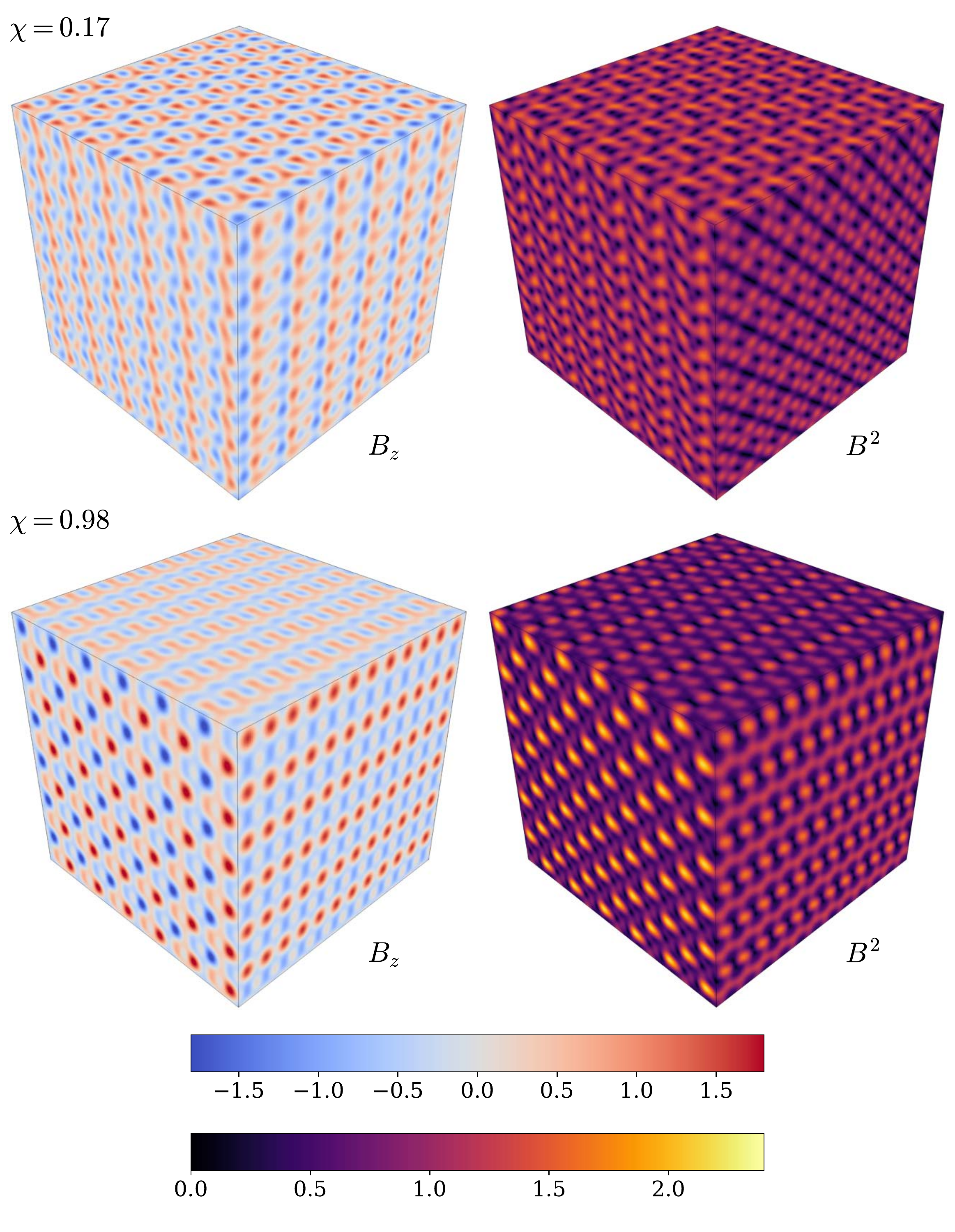}
  \caption{Magnetic-field structure of two force-free equilibria resulting from the numerical optimisation procedure described in Section \ref{simulating}. The upper colour bar describes the plots of $B_z$, while the lower colour bar describes the plots of $B^2$, in units of $\langle v_A^2 \rangle^{1/2}$ and $\langle v_A^2 \rangle$ respectively. Note that the configuration with $\chi = 0.17$ has smaller variation in the magnetic field strength than the configuration with $\chi = 0.98$. Indeed, the $\chi=0.98$ configuration is hardly `tangled' at all; it more closely resembles a collection of magnetic `blobs' (cf. the discussion in Section \ref{neglectSS}).} 
\label{fig:field_lines}
\end{figure}

\begin{figure}
  \centering
  \includegraphics[scale=0.35]{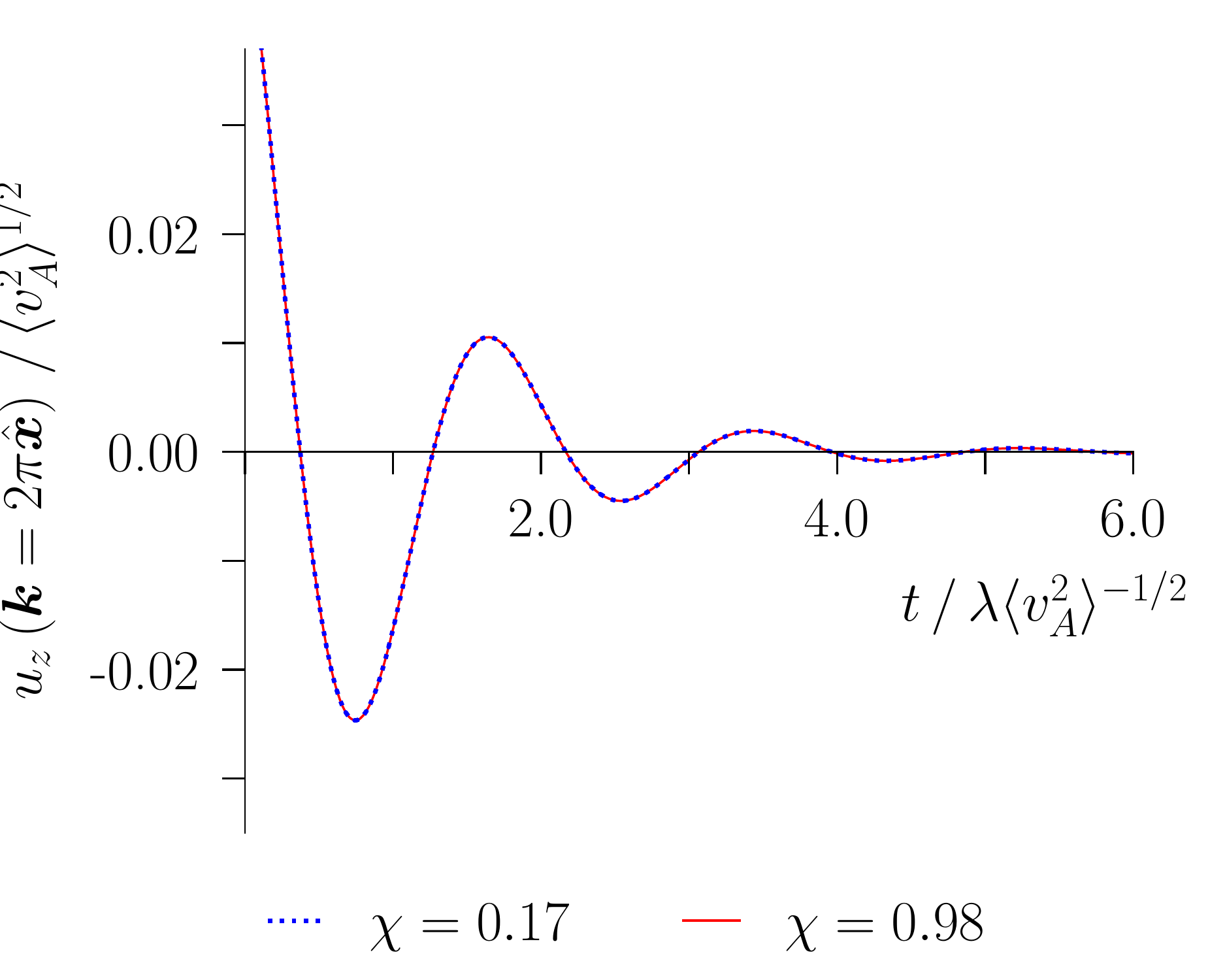}
  \caption{Wave amplitude against time for the field configurations with $\chi=0.17$ and $\chi=0.98$ (see Figure \ref{fig:field_lines} for the structure of these fields) and using sixth-order hyperviscosity with $\nu_6 \kw^6=(0.048 \lambda \langle v_A^2 \rangle^{1/2} ) \kw^2$. Despite the difference in the magnetic-field configuration, both fields show an almost identical response to the imposed wave perturbation and evolve according to \eqref{hom_WE}, as anticipated by the theory in Section \ref{neglectSS}.} 
\label{fig:hyperviscous_fig}
\vspace*{\floatsep}
  \centering
  \includegraphics[scale=0.35]{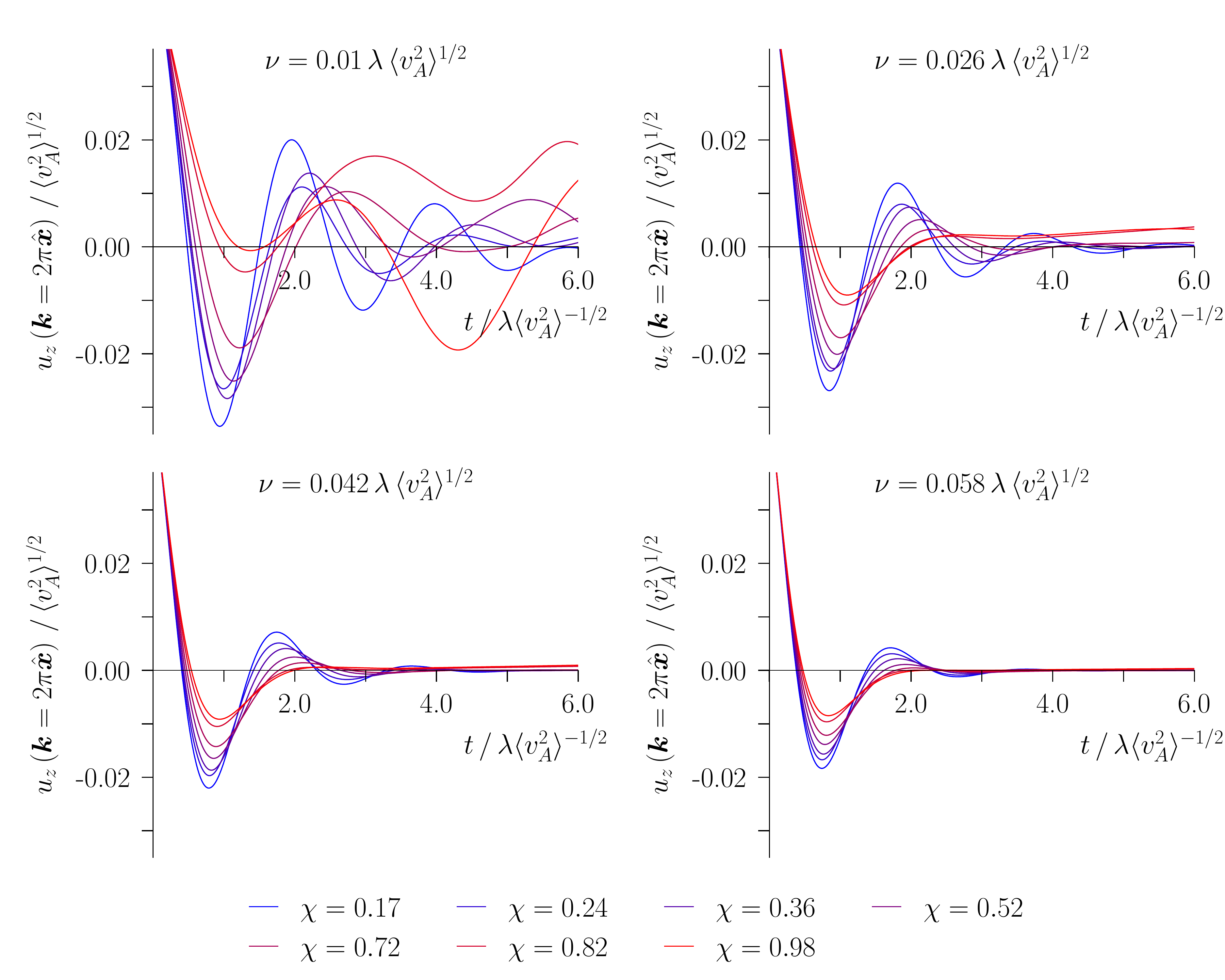}
  \caption{Wave amplitude against time for a range of values of $\chi$, plotted for different viscosities. At large viscosity, fields with larger values of $\chi$ support waves with larger period and increased damping, in accordance with Section \ref{sec_viscosity}. At small viscosities, the evolution is dominated by the onset of the instability of the equilibrium field (see Appendix \ref{App:FFinstab}).}
\label{fig:chis_w_viscosity}
\end{figure}

In order to test the analytic predictions derived in the previous sections, we conduct simulations of standing magnetoelastic waves on periodic, linear, force-free magnetic-field equilibria with different values of the parameter $\chi$. 

The linear force-free magnetic field condition is $\boldsymbol{\nabla}\times\boldsymbol{B}=\alpha \boldsymbol{B}$, where $\alpha$ is a constant. Taking the curl and using $\boldsymbol{\nabla}\bcdot\boldsymbol{B}=0$, we get a Helmholtz equation for the magnetic field, $\nabla^2\boldsymbol{B}=-\alpha^2 \boldsymbol{B}$, which shows that the modes in the Fourier expansion of $\boldsymbol{B}$ must all have $\left|\boldsymbol{k}\right|=\alpha$. The linear force-free equilibrium must then have $i \boldsymbol{k}\times\boldsymbol{B_k} = \alpha \boldsymbol{B_k}$, or
\begin{equation}
    \hat{\boldsymbol{k}}\times \text{Re}\left(\boldsymbol{B_k}\right)= \text{Im}\left(\boldsymbol{B_k}\right). \label{forcefree}
\end{equation}

To generate tangled magnetic equilibria, we take $\boldsymbol{k}=2\pi\left(12, 6, 0\right)^\mathrm{T}$, together with all its permutations and negations. This means that the magnetic-field structure is periodic on a scale of $1/6$ of the box size, which ensures the scale separation between the box scale and the scale of the tangled field. We then generate real vectors $\text{Re}\left(\boldsymbol{B_k}\right)$ subject to the condition $\text{Re}\left(\boldsymbol{B_k}\right)= \text{Re}\left(\boldsymbol{B}_{-\boldsymbol{k}}\right)$, so that $\boldsymbol{B}$ is real, and to \eqref{forcefree}, so that it is force-free, using a numerical optimisation procedure to ensure $\overline{M}_{0ij}\propto\delta_{ij}$ and to produce fields with different values of the parameter $\chi$\footnote{This process of generating magnetic equilibria is not guaranteed to produce fields that are  statistically isotropic, despite satisfying  $\overline{M}_{0ij}\propto\delta_{ij}$. As a consequence, we calculate $\chi$ directly from the tensor $R$ defined in Section \ref{sec_fosa}, rather from the isotropic expression \eqref{chi}. Anisotropy on small-scales could in principle cause large-scale modes with the same $\boldsymbol{k}$ but different spatial directions to couple. However, for all configurations that we have tested the off-diagonal terms of the matrix $\sum_{\bkp} A_{ip}(\bk, \bkp)A_{pr}(\bkp, \bk)$ are small (for the fields presented in Figure \ref{fig:field_lines}, they are small compared to the diagonal terms by a factor of $2\times10^{-3}$ for $\chi = 0.17$ and $3\times10^{-4}$ for $\chi=0.98$). We have also checked that in our simulations, negligible energy is transferred into any large-scale mode other than the originally perturbed one.}. The magnetic-field configurations thus obtained with $\chi=0.17$ and $\chi=0.98$ are shown in Figure \ref{fig:field_lines}. Curiously, we have been unable to generate a field with $\chi>1$ using this optimisation procedure, which is the condition for the large-scale mode to become unstable in the FOSA. 

For each equilibrium configuration, we introduce a sinusoidal velocity shear $\boldsymbol{u} = 0.05\, \langle v_A^2\rangle^{1/2}\sin{\left(2\pi x/\lambda\right)}\hat{\boldsymbol{z}}$, where $\lambda=2\pi /\kw$ is the box size and $\langle v_A^2 \rangle^{1/2}=\langle B_0^2 \rangle^{1/2}=\sqrt{3}\,\vM$ is the r.m.s. Alfv\'{e}n speed, and we measure the evolution of the amplitude of this mode in the Fourier representation of $\boldsymbol{u}$.

The 3D, incompressible MHD equations are solved using the Dedalus\footnote{Dedalus is available at http://dedalus-project.org} code \citep{Burns20} at $128^3$ resolution (padded for de-aliasing according to the 2/3 rule). We take the resistivity $\eta = 0$, and thus our simulations are only valid in describing the evolution at early times when the small-scale instability of the equilibrium field configuration has not yet developed and hence the magnetic field structure is still well-resolved.

\subsection{Results}

\subsubsection{Hyperviscous regime\label{sec:hyperviscous}}

As the simplest `control' case, we first present the results of simulations with hyperviscosity, i.e., with the viscous dissipation term in \eqref{MHDmomentum} replaced by $\nu\, \nabla^6 \boldsymbol{u}$, with the tangled magnetic field at a smaller scale than the hyperviscous cutoff scale. As shown in Section \ref{neglectSS}, the small-scale motions cannot develop in this regime, therefore we expect an imposed sinusoidal shear flow to evolve according to \eqref{hom_WE}, which might now be called the ZOSA prediction (Zeroth-Order Smoothing Approximation), independently of the small-scale structure of the magnetic field. This is indeed observed, as shown in Figure \ref{fig:hyperviscous_fig}, where the evolution of the wave amplitude is plotted for $\chi=0.17$ and $\chi=0.98$. There is a small relative deviation of $\sim 10^{-3}$ between the two curves and the prediction of the ZOSA \eqref{hom_WE} that is not visible on this plot.

\subsubsection{Effect of increasing $\chi$}

\begin{figure}
  \centering
  \includegraphics[scale=0.35]{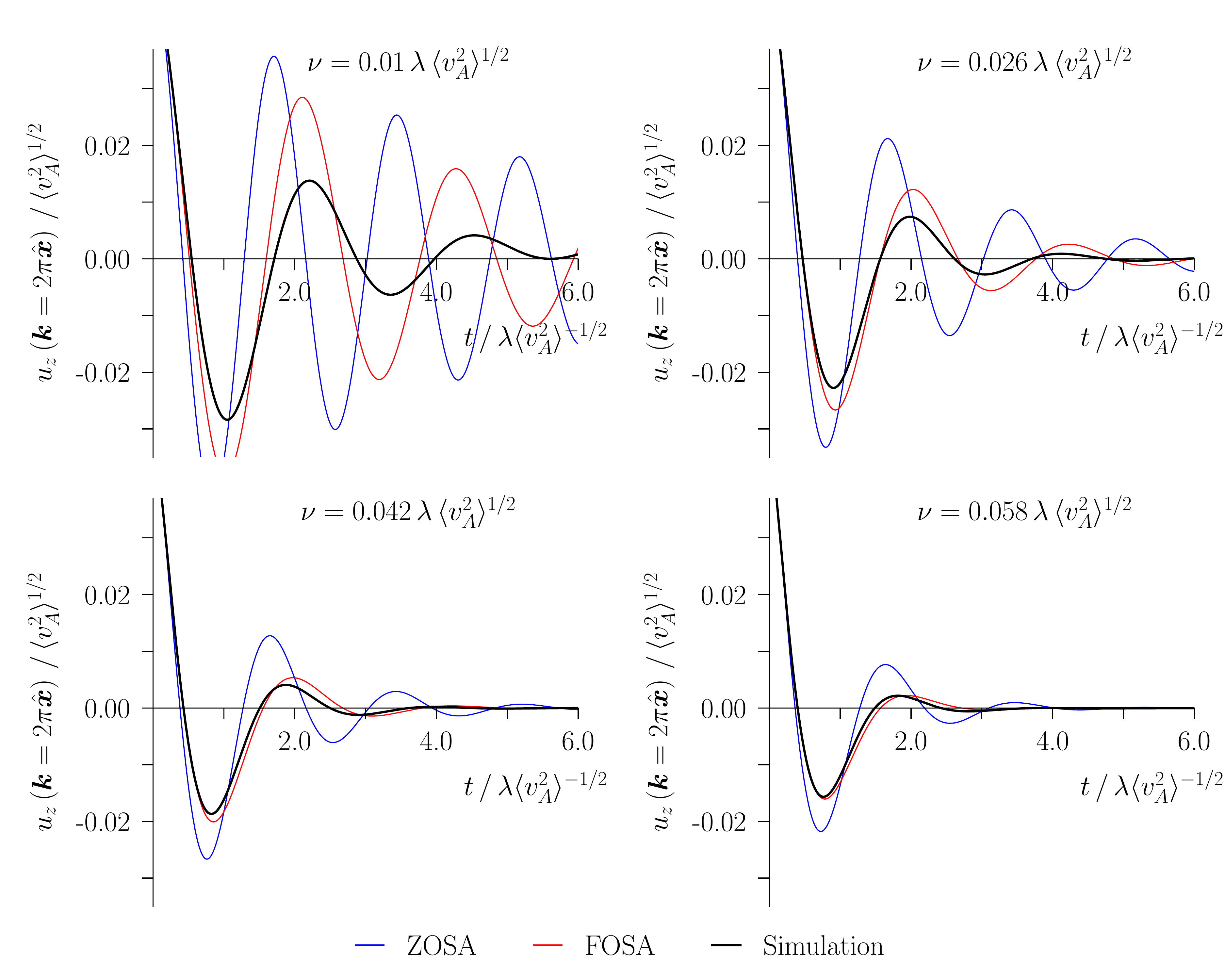}
  \caption{Wave amplitude against time for $\chi=0.51$ at a range of viscosities (black), together with the predictions of the ZOSA, (equation \ref{hom_WE}; blue), i.e., neglecting all small-scale motions, and the FOSA (equation \ref{residue}; red).}
\label{fig:viscosities}
\end{figure}

We now turn to the case of Laplacian viscosity. Figure \ref{fig:chis_w_viscosity} shows the wave evolution for a number of different values of $\chi$, at a range of Laplacian viscosities. When the viscosity is not very small, the evolution is qualitatively in agreement with the FOSA theory developed in Section \ref{sec_viscosity}, i.e., a larger $\chi$ results in a longer wave period and increased damping. At small viscosity, the evolution is dominated by the onset of the instability of the periodic force-free field, which results in a transfer of energy from small to large scales via mergers of magnetic structures (see Appendix \ref{App:NonlinInstab}) that disrupts the waves (this process is not well-resolved in our simulations). We note that the time taken for the instability to develop becomes smaller as $\chi$ increases.

\subsubsection{Comparison with the analytic theory at different viscosities}

Figure \ref{fig:viscosities} shows the evolution of the wave for $\chi = 0.51$ and for four different Laplacian viscosities, with the predictions of the ZOSA and FOSA plotted for comparison. We find that the FOSA gives a better prediction in each case, with excellent agreement for large viscosities, correctly capturing the longer wave period and the increased damping observed.

\subsubsection{Comparison with the analytic theory at different values of $\chi$}

\begin{figure}
  \centering
  \includegraphics[scale=0.35]{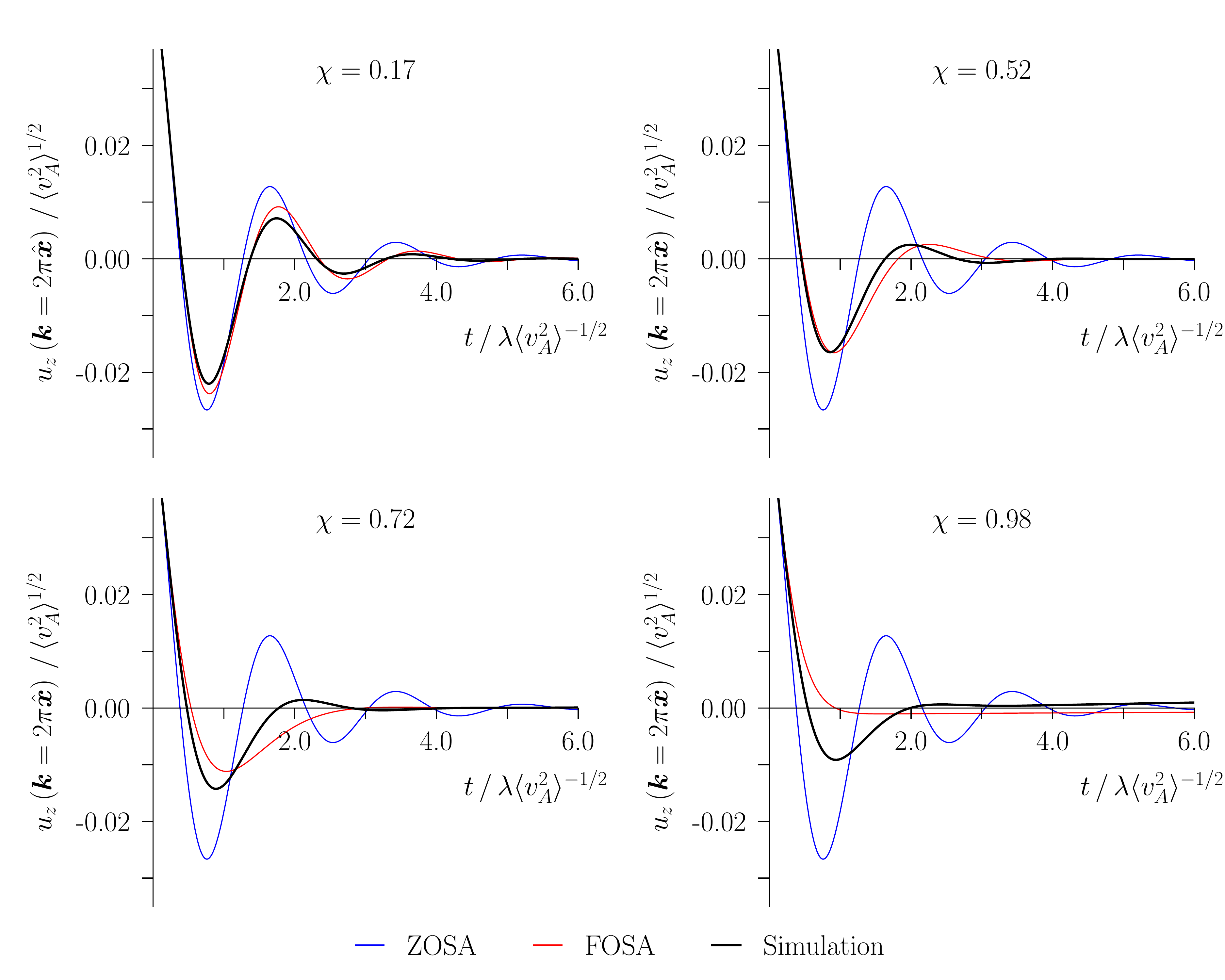}
  \caption{Wave amplitude against time for different values of $\chi$ at fixed viscosity $\nu=0.03 \lambda \langle v_A^2 \rangle^{1/2}$, together with the predictions of the ZOSA, (equation \ref{hom_WE}; blue), i.e., neglecting all small-scale motions, and the FOSA (equation \ref{residue}; red).}
\label{fig:chis_w_theory}
\end{figure}

Finally, Figure \ref{fig:chis_w_theory} shows the evolution of the wave for different values of $\chi$ at fixed viscosity, $\nu = 0.03 L \langle v_A \rangle^{1/2}$, with the predictions of the ZOSA and FOSA theories plotted for comparison. Again, we find that the FOSA gives a better prediction in each case, although at large values of $\chi$ it fails to predict the numerical result accurately.

\subsection{Discussion}\label{resultsanalysis}

Our numerical experiments show that the analytical theory presented in Section \ref{sec_viscosity} gives a good description of the dynamics of magnetoelastic waves either in the limit of small $\chi$ or of large viscosity. This is a natural result, as the FOSA assumes greatly simplified small-scale dynamics. Describing the small-scale motions precisely is most important when viscosity is small, because they are less-strongly damped, and when $\chi$ is large, because $\chi$ encodes the strength of the coupling of large and small scales.


\section{Conclusion}

In this work, we have studied the large-scale elastic dynamics of MHD equilibrium states with statistically homogeneous and isotropic tangled magnetic-field configurations. We have extended the model presented by \cite{Moffatt86} to consider how the inevitable small-scale inhomogeneity of the magnetic field structure modifies the dynamics. We have found that in the idealised case of a stable equilibrium state, the frequency of magnetoelastic waves is necessarily reduced as a result of accounting for the inhomogeneity, as a result of relaxation of small-scale structures, which reduces the elastic tension. By employing the First-Order Smoothing Approximation (FOSA), where couplings between small-scale motions are neglected, we have been able to derive a dispersion relation for magnetoelastic waves in terms of statistical properties of the magnetic-field structure. A key finding is that more intermittent fields are less elastic (see equation \ref{chiFOSA}), with the controlling parameter being the variance of the total (thermal + magnetic) pressure. By solving the initial-value problem for a large-scale pulse applied to a viscous fluid, we have further shown that the viscous damping rate of magnetoelastic waves in a magnetic tangle with small-scale inhomogeneity is greater than for a perfectly homogeneous and isotropic Maxwell stress, because the small-scale motions are themselves subject to viscous damping. As we found in Section \ref{neglectSS}, these effects are not present in a hyperviscous fluid because dynamically significant motions on the scale of the tangle are prevented.



This work is an important first step towards our ultimate goal of understanding the dynamic effect of tangled magnetic field on large-scale motions. In the future, we plan to investigate how the conclusions reached here apply to more general field configurations and to MHD turbulence. The outstanding questions that we hope to address include:

\begin{itemize}
    \item How restrictive is the equilibrium assumption? In this work, we have frequently utilised the Hermitian nature of the force operator (see Section \ref{sec:normalmodes}), which relies on the unperturbed field being an equilibrium state. While turbulent magnetic fields may be far from equilibrium, simulations of isotropic MHD turbulence have shown an excess of magnetic energy (`residual energy') at large scales \citep{Muller04,Muller05}, hinting that at these scales, the field may organise itself into quasi-equilibrium structures.
    \item How restrictive is assuming that the magnetic field has no structure at large scales? In writing \eqref{vE}, we assumed that the equilibrium Maxwell stress $M_0$ had no component at the scale of the magnetoelastic-wave motions. Furthermore, assuming that different large-scale modes are not coupled via their individual coupling to small-scale modes required that positive integer powers of $M_0$ also did not have structure at large scales (see the discussion in Section \ref{sec:normalmodes}). These assumptions are justified for synthetic fields of the sort investigated numerically in Section \ref{numerics}, but may not be realistic for MHD turbulence (see, for example, \citealt{Schekochihin04}).
    \item What is the effect of magnetic reconnection? In this work, we have exclusively considered flux-frozen magnetic fields. Indeed, as discussed in Section \ref{homogeneous_tangle}, the MHD equations cannot be written as closed evolution equations for the Maxwell stress when resistive effects are included. Intuitively, reconnection of magnetic field lines would reduce the magnetic tension that can be maintained at large scales, so we expect reconnecting magnetic fields to be less elastic.
\end{itemize}

\section*{Acknowledgements}

It is a pleasure to acknowledge useful conversations with Santiago Benavides and François Rincon. DNH was supported by an STFC studentship. The work of AAS was supported in part by the UK EPSRC grant EP/R034737/1. SAB was supported in part by STFC grant ST/S000488/1 and the Hintze Family Charitable Foundation. Simulations were performed on the Oxford Hydra cluster.

\appendix

\section{Ideal instability of linear force-free fields \label{App:FFinstab}}
\subsection{Existence of an ideal instability \label{App:FFinstabExistence}}

Force-free magnetic fields are an important class of magnetostatic equilibria characterised by a vanishing Lorentz force. Such fields satisfy the equation
\begin{equation}
    \boldsymbol{\nabla}\times \boldsymbol{B} = \alpha \boldsymbol{B}, \label{FFcond}
\end{equation}where $\alpha\left(\boldsymbol{r}\right)$ is a scalar function of position, constrained to satisfy
\begin{equation}
    \boldsymbol{B}\bcdot \boldsymbol{\nabla}\alpha = 0,
\end{equation}by the divergence-free nature of $\boldsymbol{B}$. Force-free magnetic fields have long enjoyed prominence in plasma astrophysics as a natural relaxed state for magnetically dominated systems. The first such suggestion appears to be by \cite{Lust54}, in an effort to explain early investigations that indicated pressure gradients and gravity might be insufficient to balance strong fields in stellar media.

Special significance of \emph{linear} force-free (LFF) fields, i.e., those for which $\alpha = \mathrm{constant}$, was first recognised by \cite{Chandrasekhar58}, who showed that LFF fields are among the fields that minimise the total Joule heating subject to fixed magnetic energy, and argued on thermodynamic grounds that magnetically-dominated systems should relax to this minimum-dissipation state. This argument was refined somewhat by \cite{Woltjer58a}, who showed that ideal MHD evolution conserves a quantity that later became known as the magnetic helicity,
\begin{equation}
    H = \int \mathrm{d}^3 \boldsymbol{r}\,\boldsymbol{A}\bcdot \boldsymbol{B}, \label{helicity}
\end{equation} and further showed that extremising the magnetic energy subject to the constraint of fixed magnetic helicity produced LFF fields. This \emph{extremisation} was interpreted as a \emph{minimisation} of the magnetic energy, and therefore a proof of the stability of LFF equilibria against ideal (helicity-conserving) perturbations. In a later paper, \cite{Woltjer59} argued that this conclusion had been incorrect, because the extremisation process was not guaranteed to produce minima, only extrema. Nonetheless, one state was guaranteed to be stable, the global minimum of the magnetic energy subject to fixed helicity. Substitution of \eqref{FFcond} into \eqref{helicity} gives
\begin{equation}
    H = \frac{1}{\alpha}\int \mathrm{d}^3 \boldsymbol{r}\,B^2 = \frac{2E_M}{\alpha},
\end{equation} and hence the magnetic energy is $E_M = \alpha H /2$. The force-free state with the smallest magnetic energy for a given magnetic helicity is therefore the one with the smallest value of $\alpha$ consistent with the boundary conditions; this state is guaranteed to be stable\footnote{{These early arguments were later refined by \cite{Taylor74}, who provided justification that conservation of magnetic helicity was indeed the correct constraint under which MHD fluids should relax. Taylor argued that the magnetic-field-topology-preserving nature of the MHD equations meant that any topological invariant of the magnetic field should be conserved during the evolution. This condition can be expressed as the conservation of $H_V= \int_V \mathrm{d}^3 \boldsymbol{r}\,\boldsymbol{A}\bcdot\boldsymbol{B}$ where $V$ is any flux tube; physically, this integral is the total (signed) flux linked by the flux tube $V$. However, when even small topology breaking terms are introduced to the induction equation, $H_V$ will no longer be conserved by the MHD evolution. The sum $H_{V_1}+H_{V_2}$ is nonetheless conserved when the tubes $V_1$ and $V_2$ `unlink', and hence the sum $\sum_i H_{V_i}$ over all flux tubes will still be a conserved quantity, as long as the evolution can be taken to be ideal outside of these unlinking events. The quantity $\sum_i H_{V_i}$ is just the total helicity, which, \citeauthor{Taylor74} argued, justifies its priviledged role as the only topological invariant conserved in MHD relaxation.}}.

While the question of the stability of states with larger values of $\alpha$ is not resolved by the extremisation argument, theorems have been proposed regarding the stability of such states. These theorems mostly rely on the energy principle of \cite{Bernstein58}, which states that an MHD equilibrium state is stable if the second-order change in the total energy \begin{equation}
    \delta W_2 = \frac{1}{2} \int \mathrm{d}^3 \boldsymbol{r} \left[\left(\boldsymbol{\xi}\bcdot\boldsymbol{\nabla}p_0\right)\boldsymbol{\nabla}\bcdot\boldsymbol{\xi}+\gamma p_0 \left(\boldsymbol{\nabla}\bcdot\boldsymbol{\xi}\right)^2 + \left(\boldsymbol{\nabla}\times\boldsymbol{B}_0\right)\bcdot\left(\boldsymbol{\xi}\times\delta\boldsymbol{B}\right)+\left|\delta \boldsymbol{B}\right|^2\right]\label{energyprinciple}
\end{equation} is positive for any displacement field $\boldsymbol{\xi}$, where $\gamma$ is the adiabatic index, $p$ is the thermal pressure, the subscript zero refers to equibrium quantities, and $\delta \boldsymbol{B} = \boldsymbol{\nabla}\times\left(\boldsymbol{\xi} \times \boldsymbol{B}_0\right)$. The first term is zero for a force-free equilibrium, while the second is positive definite, but can always be made small by taking $p_0 \rightarrow 0$. Hence, the only terms relevant to the stability of force-free fields are the final two. Substituting the force free condition, \eqref{FFcond}, into \eqref{energyprinciple}, we obtain
\begin{equation}
    \delta W_2 = \frac{1}{2} \int \mathrm{d}^3 \boldsymbol{r} \left[\left(\boldsymbol{\nabla}\times\delta\boldsymbol{ \boldsymbol{A}}\right)^2-\alpha\, \delta \boldsymbol{A}\bcdot \left(\boldsymbol{\nabla}\times\delta\boldsymbol{A}\right) \right], \label{FFenergyprinciple}
\end{equation}where $\delta\boldsymbol{A} = \boldsymbol{\xi}\times\boldsymbol{B_0}$. From this or equivalent expressions, it has been shown that more-or-less restricted classes of perturbations will give $\delta W_2>0$ for more-or-less restricted classes of LFF fields, and hence those equilibria are stable to such perturbations\footnote{This language, common in the literature, is a little imprecise, because a field perturbed by a displacement field $\boldsymbol{\xi}$ for which $\delta W_2 > 0$ will not necessarily tend to return to its original equilibrium state, unless the field is indeed stable to all perturbations.}. Classes of perturbations to which LFF fields have been shown to be stable (for any constant $\alpha$) include: radial expansions; displacements along one spatial direction (i.e., $\boldsymbol{\xi}=\xi\left( \boldsymbol{r}\right)\boldsymbol{n}$, where $\xi\left(\boldsymbol{r}\right)$ is an arbitrary scalar function of position and $\boldsymbol{n}$ is a constant vector); axisymmetric perturbations (of axisymmetric LFF fields only); and perturbations that vanish outside of a region with spatial extent $d<1/|\alpha|$ \citep{Woltjer58a, Molodensky74}. To each of these perturbations it is possible to add any component along $\boldsymbol{B}_0$, because this does not change $\delta \boldsymbol{A}$. 

\cite{Voslamber62}, however, showed by means of a counterexample that not all LFF fields are stable; they found that there exist axisymmetric field configurations that are unstable to a class of non-axisymmetric perturbations. This dashed the hopes of proving a general stability theorem for LFF fields until the problem was revisited by \cite{Moffatt86}, who showed by expanding $\boldsymbol{B}_0$ and $\boldsymbol{\xi}$ in Fourier modes that arbitrary periodic LFF equilibria are stable to arbitrary periodic perturbations. Indeed, the stability of LFF fields was the motivation for the consideration of the magnetoelastic wave problem by \cite{Moffatt86}. Despite the apparent conflict between  \citeauthor{Voslamber62}'s result and \citeauthor{Moffatt86}'s stability theorem, it was not until much more recently that a counterexample was explicitly presented by \cite{Er-Riani14}, who showed that there exist periodic force-free fields that are unstable to ideal periodic perturbations, as long as these perturbations are allowed to have wavevectors smaller than $\alpha$\footnote{This is to say, a periodic LFF field with periodicity $2\pi$ ($\alpha=2\pi$), in the 3-torus (or periodic box) with periodicity $2\pi$ \emph{will} be stable. This is guaranteed to be so because $2\pi$ is the smallest value of $\alpha$ consistent with the periodicity of the domain, and hence the field must be stable by the variational result of \cite{Woltjer58a}.}. Soon after, \cite{East15} were able to find energy-decreasing peturbations for a number of LFF fields by numerical minimisation of \eqref{energyprinciple} under variation of the Fourier coefficients in the expansion of $\boldsymbol{\xi}$. Indeed, they reported that they were able to find energy-decreasing peturbations for \emph{every} LFF field that they considered, when Fourier modes with wavenumbers smaller than $\alpha$ were allowed. We too have encountered instability in each of the fields that we have investigated numerically in this work.

We note that there \emph{do} exist stable periodic LFF fields, though only in the restricted case of no magnetic curvature. An example is the field $\boldsymbol{B} = B_0 \left(0,\sin\alpha x,\cos\alpha x\right)$ which represents a uniform magnetic field in any plane of constant $x$, with direction rotating as $x$ is varied. The stability of this configuration against a restricted class of periodic perturbations was demonstrated by \cite{Vekshtein89}; here, we note that for a general periodic perturbation $\boldsymbol{\xi}\left(\boldsymbol{r}\right)=\left(\xi_x\left(\boldsymbol{r}\right),\, \xi_y\left(\boldsymbol{r}\right),\, \xi_z\left(\boldsymbol{r}\right)\right)^\mathrm{T}$ it is possible to show via an elementary (though somewhat tedious) calculation that

\begin{align}
    \delta W_2 & = \frac{1}{2} B_0^2 \int \mathrm{d}^3 \boldsymbol{r} \left[\left(\cos\alpha x  \,\frac{\p \xi_x}{\p x}+\cos\alpha x \,\frac{\p \xi_y}{\p y}-\sin\alpha x \,\frac{\p \xi_z}{\p y} \right)^2 \right.\nonumber \\
    & \left.+\left(\sin\alpha x  \,\frac{\p \xi_x}{\p x}+\sin\alpha x \,\frac{\p \xi_z}{\p z}-\cos\alpha x \,\frac{\p \xi_y}{\p z} \right)^2+
    \left(\sin\alpha x  \,\frac{\p \xi_x}{\p y}+\cos\alpha x \,\frac{\p \xi_x}{\p z} \right)^2\right].
\end{align}The integrand is manifestly positive definite, so this configuration is stable to ideal perturbations, for any $\alpha$\footnote{In fact, the same result is valid for $\alpha \,x\rightarrow\Phi\left(x\right)$, $\Phi'\left(x\right)\rightarrow\alpha\left(x\right)$, so a nonlinear force-free configuration corresponding to a non-constant rate of rotation of the magnetic field direction is also stable.}.

Whether there exist any periodic LFF configurations with magnetic curvature that are stable to arbitrary perturbations is, at present, unknown, as there have been no general theorems of instability (see \citealt{Zrake16} for speculations that stable structures may exist in the \emph{nonlinear} force-free case). The existence of stable force-free configurations would highly significant for astrophysical systems where magnetic-field structures are smaller than the system size, such as the hot, rarefied plasma between galaxies in clusters.

\subsection{Physical nature of the instability \label{App:PhysNatureInstab}}

Since its (re)discovery in 2014, the instability of LFF fields has attracted attention as a means of studying particle acceleration by magnetic reconnection in a more realistic setting than the typical Harris-type current sheet configuration\footnote{We stress that the instability is ideal in nature, but current sheets naturally form in its nonlinear evolution.} \citep{Nalewajko16, Lyutikov17a}.

\cite{Lyutikov17a} have proposed a mechanism for the instability of two-dimensional LFF fields, by arguing that these fields describe a regular array (in the $xy$ plane) of alternating currents (directed along $z$). Since like currents attract, this configuration is unstable, with similarly-directed currents ultimately merging.

\begin{figure}
  \centering
  \includegraphics[scale=0.35]{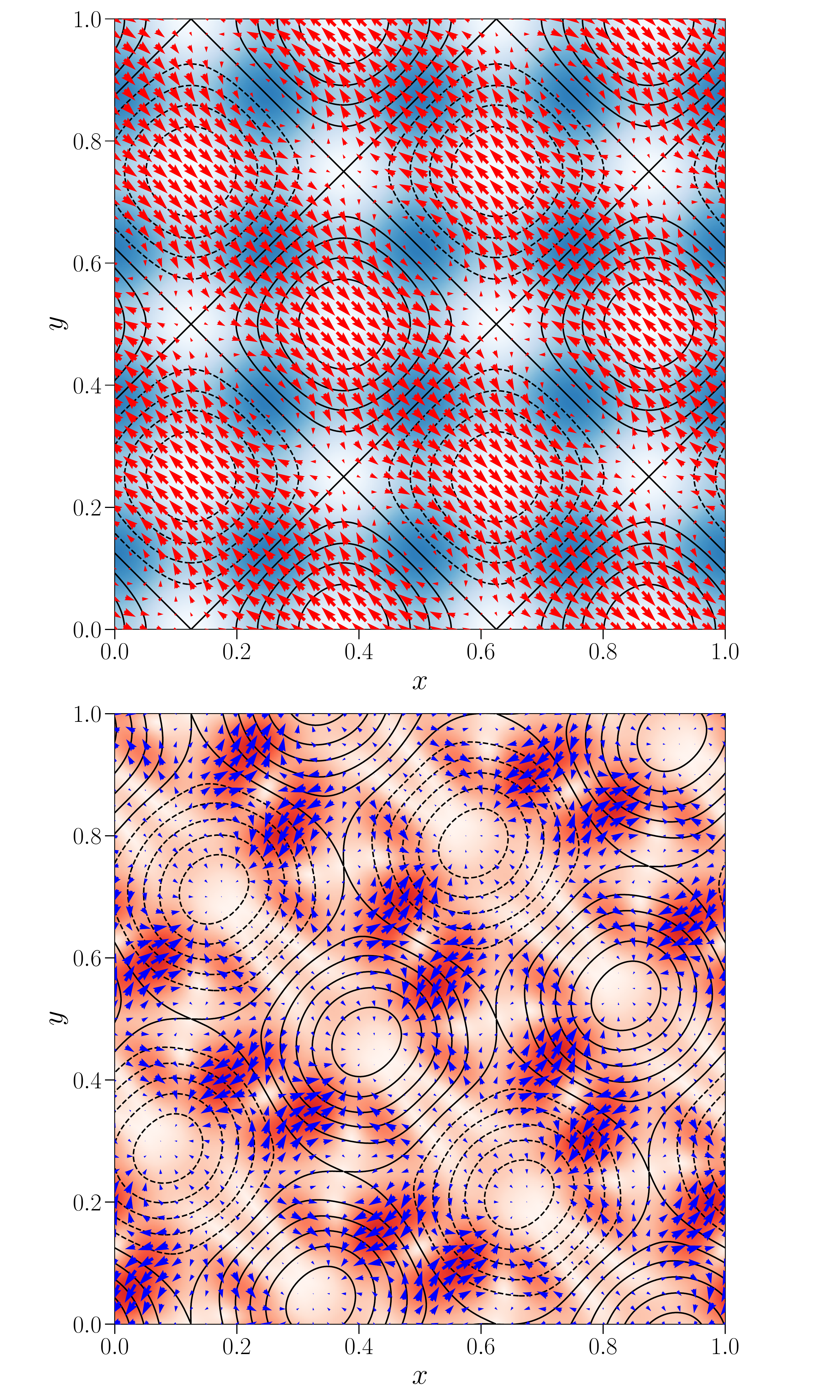}
  \caption{Upper panel: Black lines are level lines of $B_z$ for the field defined by \eqref{mechanismfield}. By the LFF condition and $z$-independent nature of the field, these are also the field lines of the dynamically-important $xy$-plane field. Solid black lines have $B_z>0$, so the in-plane field has anticlockwise circulation, while dashed black lines have $B_z<0$, so the in-plane field has clockwise circulation. The strength of the in-plane field is shown in blue. Red arrows show a perturbation for which $\delta W <0$. Lower panel: The field configuration obtained by applying this perturbation. Black lines again show the $xy$-plane field, while blue arrows show the non-compressive part of the magnetic tension force, with the size of this force shown in red.}
\label{fig:mechanism}
\end{figure}

In this work, we have been primarily motivated by the manifestation of the Lorentz force as a local magnetic tension in incompressible MHD. It is therefore instructive to see how the instability arises in this picture. Figure \ref{fig:mechanism} shows the ideal instability of the 2D LFF given by
\begin{equation}
    \boldsymbol{B}=B_0 \left(-\sin4\pi y, \; \cos 4\pi x, \;\cos4\pi y - \sin 4 \pi x \right),\label{mechanismfield}
\end{equation} which has $\alpha = 4\pi$. Since the field is translationally invariant in $z$, only its component in the $xy$ plane contributes to the magnetic tension. An unstable perturbation can be obtained by trialling a truncated Fourier series in \eqref{energyprinciple}; evaluation of the integral gives a quadratic form in the coefficients of the Fourier modes, which can be diagonalised and the perturbation that minimises $\delta W$ obtained. By carrying out this proceedure, we find that $\xi_x = 7.34754 \cos (2 \pi  x+2 \pi y)-7.34754 \sin (2 \pi  x+2 \pi y))+1.30549
   \cos (2 \pi  x-6 \pi y))-1.30549 \sin (2 \pi  x-6 \pi y))+0.447807 \cos (6 \pi x-2 \pi y))+0.447807 \sin (6 \pi x-2 \pi y))+ \sin (6 \pi x+6 \pi y))+\cos (6 \pi  x+6 \pi y)),\quad\xi_y = -7.34754 \cos (2 \pi  x+2 \pi  y)+7.34754 \sin (2 \pi  x+2 \pi  y)+0.435165 \cos (2 \pi  x-6 \pi  y)-0.435165
   \sin (2 \pi  x-6 \pi  y)+1.34342 \cos (6 \pi  x-2 \pi  y)+1.34342 \sin (6 \pi  x-2 \pi  y)- \sin (6 \pi  x+6 \pi 
   y)- \cos (6 \pi 
   x+6 \pi  y$) gives $\delta W_2<0$. This perturbation is shown in Figure \ref{fig:mechanism}, together with the perturbed magnetic-field structure. We observe that the perturbation represents a non-trivial deformation of the magnetic field structure that mostly consists of the ``magnetic cells'' that make up the field configuration sliding past each other in layers, with smaller perturbations to the field at the boundaries of each cell that generate a magnetic tension force whose net direction is along the direction of the displacement of the layer, preventing the configuration from returning to its original state.
   
The timescale for the development of instability can be estimated analytically using energy conservation for the truncated system:
\begin{equation}
    \frac{\dd}{\dd t}\left(\frac{1}{2}\int \dd^3\boldsymbol{r}\,\dot{\xi}^2 + \delta W\right) = -\nu\int \dd^3\boldsymbol{r}|\boldsymbol{\nabla} \boldsymbol{\xi}|^2.
\end{equation}When the fluid is inviscid, the timescale can be estimated by balancing the kinetic and potential energy terms and solving the resulting eigenvalue problem for the growth rate of the unstable mode. For a viscously-dominated fluid, the rate of change of potential energy is balanced with the rate of energy dissipation by viscosity.

Solving the eigenvalue problem for the truncated system, we find growth rates of $0.0930196\,\alpha B_0$ for the inviscid case, and $0.0135899\, B_0^2/\nu$ for the viscous case. The dimensional form of these results is inevitable, but the numerical prefactors turn out to be small, particularly in the viscous case, indicating that the growth rate of the instability should be slower than a na\"{i}ve estimate of the kind made in Section \ref{subsec:instabintro} would indicate. The reason for this is apparent from Figure \ref{fig:mechanism}, which shows that force responsible for driving the relative motions of the layers of magnetic cells is mostly generated by the perturbations to the field at the boundary of each cell, and the component of this force along the direction of motion of the layer is small. The viscous growth rate is further suppressed by the fact that the perturbations at the boundaries of each cell that are responsible for generating forces are at a smaller scale than the cell size, and hence are damped more strongly by viscosity.

Similar considerations may apply to the more complicated 3D `tangled' equilibria that we have considered in the main text, explaining why the small-scale instability develops slowly in our simulations. \cite{Lyutikov17a} have also argued that 3D magnetic configurations should be \emph{more} stable than 2D ones because the field cannot be naturally decomposed into sliding layers.

\subsection{Nonlinear evolution \label{App:NonlinInstab}}

The nonlinear evolution of the instability has been extensively studied using both force-free electrodynamics (FFE) simulations \citep{East15,Zrake16,Lyutikov17a} and kinetic simulations \citep{Nalewajko16,Yuan16,Lyutikov17a,Nalewajko18} of simple LFF configurations, the so-called Arnold-Beltrami-Childress (ABC) fields, of which \eqref{mechanismfield} is an example.

In these studies, the late-time evolution is characterised by an ``untangling'' and merging of magnetic flux tubes, with an associated transfer of magnetic energy from small to large scales. Indeed, \cite{Blandford17} suggested that the particle acceleration that occurs during such an untangling may power dramatic flares in high energy astrophysical sources. The ultimate state of the system is the stable LFF state with the smallest $\alpha$ compatible with the size of the box, in accordance with JB Taylor Relaxation (see Section \ref{App:FFinstabExistence}). Interestingly, in 2D, \cite{Zrake16} have found with FFE simulations that the system does not relax to a LFF state, but instead to a configuration of \emph{nonlinear} force-free magnetic `bubbles', which they argue is a consequence of the existence in 2D of invariants additional to magnetic helicity. \cite{East15} found indications that it is possible for the nonlinear system to evolve into a transient LFF state with $\alpha$ larger than the smallest value permitted by the geometry, though this does not appear to have been reproduced in later studies.

\section{Proof that the amplitudes of fast Laplace modes of the large-scale motion vanish as $\epsilon\rightarrow0$ \label{App:LaplaceAmpl}}

In this appendix, we show that the Laplace amplitudes of any fast motion on large scales vanishes compared to the amplitude of slow motions, as $\epsilon\rightarrow0$. This justifies the neglect of $p^2$ in the denominator of the coupling term in \eqref{A10}.

Retaining $p^2$ in \eqref{A10}, the dispersion relation \eqref{viscousdisprel} becomes 

\begin{equation}
    D(p_n) \equiv p_n^2+\vM^2k^2-\sum_{\bkp}\frac{A_{zj}(\bkw, \bkp) A_{jz}(\bkp, \bkw)}{p_n^2+k'^2 \vM^2 + \nu k'^2 p_n} +\nu k^2 p_n = 0.\label{Appdisprel}
\end{equation}Using the non-dimensionalisation $\hat{\nu}=\nu \kw/\vM$, $\hat{p}=p/\kw\vM$, we can write \eqref{Appdisprel} as

\begin{equation}
    \hat{p}^2+1-\sum_{\boldsymbol{k}'} \frac{A_{zj}(\bkw, \bkp) A_{jz}(\bkp, \bkw)/k'^2}{\epsilon_{k'}^2\hat{p}^2 +1 + \hat{\nu} \hat{p}}+\hat{\nu}\hat{p} = 0. \label{D=0_app}
\end{equation}where $\epsilon_{k'}=k/k'\sim \epsilon$, and, as before, $\epsilon \sim \kw/\kt \ll 1$.This equation has $2W+2$ solutions, where $W$ is the number of distinct wavenumbers among the small-scale modes to which the large-scale perturbation couples -- this corresponds to the number of distinct small-scale frequencies that are possible\footnote{$W\neq N$ if the large-scale perturbation couples to both of the small-scale modes $\boldsymbol{\xi}_i(\boldsymbol{k})$ and $\boldsymbol{\xi}_j(\boldsymbol{k}')$ where $|\bk|=|\bkp|$ but $\bk\neq\bkp$ or $i\neq j$.}. To evaluate the Laplace amplitude according to \eqref{residue}, it is helpful to factorise the dispersion relation. Factoring out $1/(p^2+1/\epsilon_{k'}^2+\hat{\nu}\hat{p}/\epsilon_{k'}^2)$ for each of the $W$ terms in the sum in \eqref{D=0_app} leaves a polynomial whose leading coefficient is $1$, and whose roots are all the solutions of \eqref{D=0_app}: this polynomial can be factorised as $\prod_{m}\left(\hat{p}-\hat{p}_{m}\right)$, i.e.,
\begin{align}
    \hat{p}^2+1-\sum_{\boldsymbol{k}'} \frac{A_{zj}(\bkw, \bkp) A_{jz}(\bkp, \bkw)/k'^2}{\epsilon_{k'}^2\hat{p}^2 +1 + \hat{\nu} \hat{p}}+\hat{\nu}\hat{p}=\prod_{k'}\frac{1}{p^2+1/\epsilon_{k'}^2+\hat{\nu}\hat{p}/\epsilon_{k'}^2} \prod_{m=1}^{2W+2}\left(p-p_m\right), \label{D=02}
\end{align}where the first product is over the $W$ distinct values of $k'=|\bkp|$.

According to \eqref{residue}, the Laplace amplitudes are computed according to
\begin{align}
    A_n& = \text{Res} \left[ \frac{p}{D(p
 )} , p\rightarrow p_n \right]\nonumber\\
 &=\lim_{\hat{p}\rightarrow \hat{p}_n}\frac{\displaystyle\hat{p}\left(\hat{p}-\hat{p}_n\right)\prod_{k'}\left(\hat{p}^2+1/\epsilon_{k'}^2+\hat{\nu}\hat{p}/\epsilon_{k'}^2\right)}{\displaystyle\prod_{m=1}^{2W+2}\left(\hat{p}-\hat{p}_m\right)}, \nonumber \\
    & = \frac{\displaystyle\hat{p}_n\prod_{k'}\left(\hat{p}^2+1/\epsilon_{k'}^2+\hat{\nu}\hat{p}/\epsilon_{k'}^2\right)}{\displaystyle\prod_{m=1,\,m\neq n}^{2W+2}\left(\hat{p}_n-\hat{p}_m\right)}. \label{amplformula}
\end{align}The scaling of $A_n$ with $\epsilon$ will depend on the ordering of $\hat{\nu}$ with respect to $\epsilon$ -- this determines whether or not the small-scale magnetic field is viscously dominated. Generally, we can write $\hat{\nu}=O\left(\epsilon^\beta\right)$ and consider the scaling of $A_n$ for different values of $\beta$.

If $\beta\geq1$, the small-scale field is not overdamped by viscosity. Any solution for which $\epsilon_{k'}^2\hat{p}^2$ is not negligible in the denominator of \eqref{D=0_app} is a balance of $\epsilon_{k'}^2\hat{p}^2$ and $1$ (and $\hat{\nu}\hat{p}$, if $\beta=1$), so $p\sim\epsilon^{-1}$. We then have
\begin{equation}
    \prod_{m\neq n}\left(p_n-p_m\right) \sim \epsilon^{-(2W+1)}.
\end{equation} Each term in the product $\prod_{k'}\left(p^2+1/\epsilon_{k'}^2+\hat{\nu}\hat{p}/\epsilon_{k'}^2\right)$ is $\sim 1/\epsilon^2$ apart from one, which is that for which $p_n$ is an approximate root. The leading-order scaling of this term will be $\epsilon^2$, in order to balance the other terms in \eqref{D=0_app}. Therefore,
\begin{equation}
\prod_{k'}\left(p^2+1/\epsilon_{k'}^2+\hat{\nu}\hat{p}/\epsilon_{k'}^2\right)\sim \epsilon^{-2(W-1)}\epsilon^2=\epsilon^{4-2W}.
\end{equation}With these scalings, \eqref{amplformula} gives $A_n\sim\epsilon^4$.

Alternatively, if $\beta<1$, the small-scale field is strongly damped by viscosity. We argue similarly to before: any solution for which $\epsilon_{k'}^2\hat{p}^2$ is not negligible in the denominator of \eqref{D=0_app} is a balance of $\epsilon_{k'}^2\hat{p}^2$ and $\hat{\nu}\hat{p}$, so $p\sim\epsilon^{\beta-2}$. We then have
\begin{equation}
    \prod_{m\neq n}\left(p_n-p_m\right) \sim \left(\epsilon^{\beta-2} \right) ^{2W+1},
\end{equation} while
\begin{equation}
    \prod_{k'}\left(p^2+1/\epsilon_{k'}^2+\hat{\nu}\hat{p}/\epsilon_{k'}^2\right)\sim \left(\epsilon^{2\beta-4} \right)^{W-1}\epsilon^{4-2\beta}\sim\left(\epsilon^{\beta-2}\right)^{2W-4}.
\end{equation}With these scalings, \eqref{amplformula} gives $A_n\sim\epsilon^{8-4\beta}$.

In both cases, $A_n\rightarrow 0$ as $\epsilon\rightarrow 0$.





\bibliographystyle{jpp}

\bibliography{elasticityI_mod}

\begin{thebibliography}{35}
\expandafter\ifx\csname natexlab\endcsname\relax\def\natexlab#1{#1}\fi
\def\au#1{#1} \def\ed#1{#1} \def\yr#1{#1}\def\at#1{#1}\def\jt#1{\textit{#1}}
  \def\bt#1{#1}\def\bvol#1{\textbf{#1}} \def\vol#1{#1} \def\pg#1{#1}
  \def\publ#1{#1}\def\arxiv#1{#1}\def\org#1{#1}\def\st#1{\textit{#1}}

\bibitem[{Bernstein} {\em et~al.\/}(1958){Bernstein}, {Frieman}, {Kruskal} \&
  {Kulsrud}]{Bernstein58}
{\sc \au{{Bernstein}, I.~B.}, \au{{Frieman}, E.~A.}, \au{{Kruskal}, M.~D.} \&
  \au{{Kulsrud}, R.~M.}} \yr{1958}  \at{{An energy principle for hydromagnetic
  stability problems}}.  \jt{Proc. R. Soc. Lond.}  \bvol{244},  \pg{17}.

\bibitem[{Blandford} {\em et~al.\/}(2017){Blandford}, {Yuan}, {Hoshino} \&
  {Sironi}]{Blandford17}
{\sc \au{{Blandford}, R.}, \au{{Yuan}, Y.}, \au{{Hoshino}, M.} \& \au{{Sironi},
  L.}} \yr{2017}  \at{{Magnetoluminescence}}.  \jt{Space Sci. Rev.}
  \bvol{207},  \pg{291}.

\bibitem[{Brandenburg} \& {Subramanian}(2005)]{Brandenburg05}
{\sc \au{{Brandenburg}, A.} \& \au{{Subramanian}, K.}} \yr{2005}
  \at{{Astrophysical magnetic fields and nonlinear dynamo theory}}.  \jt{Phys.
  Rep.}  \bvol{417},  \pg{1}.

\bibitem[{Burns} {\em et~al.\/}(2020){Burns}, {Vasil}, {Oishi}, {Lecoanet} \&
  {Brown}]{Burns20}
{\sc \au{{Burns}, K.~J.}, \au{{Vasil}, G.~M.}, \au{{Oishi}, J.~S.},
  \au{{Lecoanet}, D.} \& \au{{Brown}, B.~P.}} \yr{2020}  \at{{Dedalus: A
  flexible framework for numerical simulations with spectral methods}}.
  \jt{Phys. Rev. Research}  \bvol{2},  \pg{023068}.

\bibitem[{Chandrasekhar} \& {Woltjer}(1958)]{Chandrasekhar58}
{\sc \au{{Chandrasekhar}, S.} \& \au{{Woltjer}, L.}} \yr{1958}  \at{{On
  force-free magnetic fields}}.  \jt{Proc. Natl. Acad. Sci. U.S.A.}  \bvol{44},
   \pg{285}.

\bibitem[{Chen} \& {Diamond}(2020)]{Chen20}
{\sc \au{{Chen}, C.} \& \au{{Diamond}, P.~H.}} \yr{2020}  \at{{Potential
  vorticity mixing in a tangled magnetic field}}.  \jt{Astrophys. J.}
  \bvol{892},  \pg{24}.

\bibitem[{East} {\em et~al.\/}(2015){East}, {Zrake}, {Yuan} \& {Bland
  ford}]{East15}
{\sc \au{{East}, W.~E.}, \au{{Zrake}, J.}, \au{{Yuan}, Y.} \& \au{{Bland ford},
  R.~D.}} \yr{2015}  \at{{Spontaneous decay of periodic magnetostatic
  equilibria}}.  \jt{Phys. Rev. Lett.}  \bvol{115},  \pg{095002}.

\bibitem[{Er-Riani} {\em et~al.\/}(2014){Er-Riani}, {Naji} \& {El
  Jarroudi}]{Er-Riani14}
{\sc \au{{Er-Riani}, M.}, \au{{Naji}, A.} \& \au{{El Jarroudi}, M.}} \yr{2014}
  \at{{A note on the stability of Beltrami fields for compressible fluid
  flows}}.  \jt{Int. J. Non-Linear Mech.}  \bvol{67},  \pg{231}.

\bibitem[{Gruzinov} \& {Diamond}(1996)]{Gruzinov96}
{\sc \au{{Gruzinov}, A.~V.} \& \au{{Diamond}, P.~H.}} \yr{1996}  \at{{Nonlinear
  mean field electrodynamics of turbulent dynamos}}.  \jt{Phys. Plasmas}
  \bvol{3},  \pg{1853}.

\bibitem[{Krause} \& {Raedler}(1980)]{Krause80}
{\sc \au{{Krause}, F.} \& \au{{Raedler}, K.~H.}} \yr{1980} {\em {Mean-field
  magnetohydrodynamics and dynamo theory}\/}.  \publ{Elsevier}.

\bibitem[{Kulsrud}(2005)]{Kulsrud05}
{\sc \au{{Kulsrud}, R.~M.}} \yr{2005} {\em {Plasma physics for
  astrophysics}\/}.  \publ{Princeton University Press}.

\bibitem[{L{\"u}st} \& {Schl{\"u}ter}(1954)]{Lust54}
{\sc \au{{L{\"u}st}, R.} \& \au{{Schl{\"u}ter}, A.}} \yr{1954}  \at{{Kraftfreie
  magnetfelder}}.  \jt{Z. Astrophys.}  \bvol{34},  \pg{263}.

\bibitem[{Lyutikov} {\em et~al.\/}(2017){Lyutikov}, {Sironi}, {Komissarov} \&
  {Porth}]{Lyutikov17a}
{\sc \au{{Lyutikov}, M.}, \au{{Sironi}, L.}, \au{{Komissarov}, S.~S.} \&
  \au{{Porth}, O.}} \yr{2017}  \at{{Particle acceleration in relativistic
  magnetic flux-merging events}}.  \jt{J. Plasma Phys.}  \bvol{83},
  \pg{635830602}.

\bibitem[{Maron} {\em et~al.\/}(2004){Maron}, {Cowley} \&
  {McWilliams}]{Maron04}
{\sc \au{{Maron}, J.}, \au{{Cowley}, S.} \& \au{{McWilliams}, J.}} \yr{2004}
  \at{{The nonlinear magnetic cascade}}.  \jt{Astrophys. J.}  \bvol{603},
  \pg{569}.

\bibitem[{Moffatt}(1978)]{Moffatt78}
{\sc \au{{Moffatt}, H.~K.}} \yr{1978} {\em {Magnetic field generation in
  electrically conducting fluids}\/}.  \publ{Cambridge University Press}.

\bibitem[{Moffatt}(1986)]{Moffatt86}
{\sc \au{{Moffatt}, H.~K.}} \yr{1986}  \at{{Magnetostatic equilibria and
  analogous Euler flows of arbitrarily complex topology. II - Stability
  considerations}}.  \jt{J. Fluid Mech}  \bvol{166},  \pg{359}.

\bibitem[{Molodensky}(1974)]{Molodensky74}
{\sc \au{{Molodensky}, M.~M.}} \yr{1974}  \at{{Equilibrium and stability of
  force-free magnetic field}}.  \jt{Sol. Phys.}  \bvol{39},  \pg{393}.

\bibitem[{M{\"u}ller} \& {Grappin}(2004)]{Muller04}
{\sc \au{{M{\"u}ller}, W.~C.} \& \au{{Grappin}, R.}} \yr{2004}  \at{{The
  residual energy in freely decaying magnetohydrodynamic turbulence}}.
  \jt{Plasma Phys. Control. Fusion}  \bvol{46},  \pg{B91}.

\bibitem[{M{\"u}ller} \& {Grappin}(2005)]{Muller05}
{\sc \au{{M{\"u}ller}, W.~C.} \& \au{{Grappin}, R.}} \yr{2005}  \at{{Spectral
  energy dynamics in magnetohydrodynamic turbulence}}.  \jt{Phys. Rev. Lett.}
  \bvol{95},  \pg{114502}.

\bibitem[{Nalewajko}(2018)]{Nalewajko18}
{\sc \au{{Nalewajko}, K.}} \yr{2018}  \at{{Three-dimensional kinetic
  simulations of relativistic magnetostatic equilibria}}.  \jt{Mon. Not. R.
  Astron. Soc.}  \bvol{481},  \pg{4342}.

\bibitem[{Nalewajko} {\em et~al.\/}(2016){Nalewajko}, {Zrake}, {Yuan}, {East}
  \& {Blandford}]{Nalewajko16}
{\sc \au{{Nalewajko}, K.}, \au{{Zrake}, J.}, \au{{Yuan}, Y.}, \au{{East},
  W.~E.} \& \au{{Blandford}, R.~D.}} \yr{2016}  \at{{Kinetic simulations of the
  lowest-order unstable mode of relativistic magnetostatic equilibria}}.
  \jt{Astrophys. J.}  \bvol{826},  \pg{115}.

\bibitem[{Ogilvie} \& {Proctor}(2003)]{Ogilvie03}
{\sc \au{{Ogilvie}, G.~I.} \& \au{{Proctor}, M.~R.~E.}} \yr{2003}  \at{{On the
  relation between viscoelastic and magnetohydrodynamic flows and their
  instabilities}}.  \jt{J. Fluid Mech}  \bvol{476},  \pg{389}.

\bibitem[{Qin} {\em et~al.\/}(2019){Qin}, {Salipante}, {Hudson} \&
  {Arratia}]{Qin19}
{\sc \au{{Qin}, B.}, \au{{Salipante}, P.~F.}, \au{{Hudson}, S.~D.} \&
  \au{{Arratia}, P.~E.}} \yr{2019}  \at{{Upstream vortex and elastic wave in
  the viscoelastic flow around a confined cylinder}}.  \jt{J. Fluid Mech}
  \bvol{864},  \pg{R2}.

\bibitem[{Rincon}(2019)]{Rincon19}
{\sc \au{{Rincon}, F.}} \yr{2019}  \at{{Dynamo theories}}.  \jt{J. Plasma
  Phys.}  \bvol{85},  \pg{205850401}.

\bibitem[{Schekochihin} {\em et~al.\/}(2002){Schekochihin}, {Cowley},
  {Hammett}, {Maron} \& {McWilliams}]{Schekochihin02}
{\sc \au{{Schekochihin}, A.~A.}, \au{{Cowley}, S.~C.}, \au{{Hammett}, G.~W.},
  \au{{Maron}, J.~L.} \& \au{{McWilliams}, J.~C.}} \yr{2002}  \at{{A model of
  nonlinear evolution and saturation of the turbulent MHD dynamo}}.  \jt{New J.
  Phys.}  \bvol{4},  \pg{84}.

\bibitem[{Schekochihin} {\em et~al.\/}(2004){Schekochihin}, {Cowley}, {Taylor},
  {Maron} \& {McWilliams}]{Schekochihin04}
{\sc \au{{Schekochihin}, A.~A.}, \au{{Cowley}, S.~C.}, \au{{Taylor}, S.~F.},
  \au{{Maron}, J.~L.} \& \au{{McWilliams}, J.~C.}} \yr{2004}  \at{{Simulations
  of the small-scale turbulent dynamo}}.  \jt{Astrophys. J.}  \bvol{612},
  \pg{276}.

\bibitem[{Taylor}(1974)]{Taylor74}
{\sc \au{{Taylor}, J.~B.}} \yr{1974}  \at{{Relaxation of toroidal plasma and
  generation of reverse magnetic fields}}.  \jt{Phys. Rev. Lett.}  \bvol{33},
  \pg{1139}.

\bibitem[{Vekshtein}(1989)]{Vekshtein89}
{\sc \au{{Vekshtein}, G.~E.}} \yr{1989}  \at{{Magnetohydrodynamic stability of
  force-free magnetic fields in a rarefied plasma.}}  \jt{J. Exp. Theor. Phys.}
   \bvol{96},  \pg{1263}.

\bibitem[{Voslamber} \& {Callebaut}(1962)]{Voslamber62}
{\sc \au{{Voslamber}, D.} \& \au{{Callebaut}, D.~K.}} \yr{1962}  \at{{Stability
  of force-free magnetic fields}}.  \jt{Phys. Rev.}  \bvol{128},  \pg{2016}.

\bibitem[{Williams}(2004)]{Williams04}
{\sc \au{{Williams}, P.~T.}} \yr{2004}  \at{{Turbulent magnetohydrodynamic
  elasticity: Boussinesq-like approximations for steady shear}}.  \jt{New
  Astron.}  \bvol{10},  \pg{133}.

\bibitem[{Woltjer}(1958)]{Woltjer58a}
{\sc \au{{Woltjer}, L.}} \yr{1958}  \at{{A theorem on force-free magnetic
  fields}}.  \jt{Proc. Natl. Acad. Sci. U.S.A.}  \bvol{44},  \pg{489}.

\bibitem[{Woltjer}(1959)]{Woltjer59}
{\sc \au{{Woltjer}, L.}} \yr{1959}  \at{{Hydromagnetic equilibrium II.
  Stability in the variational formulation}}.  \jt{Proc. Natl. Acad. Sci.
  U.S.A.}  \bvol{45},  \pg{769}.

\bibitem[{Yuan} {\em et~al.\/}(2016){Yuan}, {Nalewajko}, {Zrake}, {East} \&
  {Blandford}]{Yuan16}
{\sc \au{{Yuan}, Y.}, \au{{Nalewajko}, K.}, \au{{Zrake}, J.}, \au{{East},
  W.~E.} \& \au{{Blandford}, R.~D.}} \yr{2016}  \at{{Kinetic study of
  radiation-reaction-limited particle acceleration during the relaxation of
  unstable force-free equilibria}}.  \jt{Astrophys. J.}  \bvol{828},  \pg{92}.

\bibitem[{Zel'dovich} {\em et~al.\/}(1983){Zel'dovich}, {Ruzmaikin} \&
  {Sokoloff}]{Zeldovich83}
{\sc \au{{Zel'dovich}, Y.~B.}, \au{{Ruzmaikin}, A.~A.} \& \au{{Sokoloff},
  D.~D.}} \yr{1983} {\em {Magnetic fields in astrophysics}\/}.  \publ{Gordon
  and Breach}.

\bibitem[{Zrake} \& {East}(2016)]{Zrake16}
{\sc \au{{Zrake}, J.} \& \au{{East}, W.~E.}} \yr{2016}  \at{{Freely decaying
  turbulence in force-free electrodynamics}}.  \jt{Astrophys. J.}  \bvol{817},
  \pg{89}.

\end{thebibliography}

\end{document}